\documentclass[lettersize,journal]{IEEEtran}
\usepackage{amsmath,amsfonts}
\usepackage{algorithmic}
\usepackage{algorithm}
\usepackage{array}
\usepackage{tabularx}
\usepackage{textcomp}
\usepackage{stfloats}
\usepackage{url}
\usepackage{verbatim}
\usepackage{graphicx}
\usepackage{cite}
\usepackage{subfigure}  
\usepackage{float}  
\usepackage{textcomp}
\usepackage{tikz,xcolor}
\usepackage{hyperref}
\usepackage{multirow}
\usepackage{booktabs}
\makeatletter
\renewcommand{\maketag@@@}[1]{\hbox{\m@th\normalsize\normalfont#1}}%
\makeatother

\begin{document}

\title{Deep Learning Based Multi-Node ISAC 4D Environmental Reconstruction with Uplink- Downlink Cooperation}


\author{
  Bohao Lu,
  Zhiqing Wei,~\IEEEmembership{Member,~IEEE},
  Huici Wu,~\IEEEmembership{Member,~IEEE},
  Xinrui Zeng,
  Lin Wang,~\IEEEmembership{Student Member,~IEEE},
  Xi Lu,
  Dongyang Mei,
  and Zhiyong Feng,\label{key}~\IEEEmembership{Senior Member,~IEEE}

\thanks{This work was supported in part by the National
		Key Research and Development Program under Grant 2020YFA0711302,
		in part by the National Natural Science Foundation of China (NSFC) under Grant 62271081, 92267202, and U21B2014.
        
         Bohao Lu, Zhiqing Wei, Huici Wu, Lin Wang, Xi Lu, Dongyang Mei and Zhiyong Feng are with the Key Laboratory of Universal Wireless Communications, Ministry of Education, School of Information and Communication Engineering, Beijing University of Posts and Telecommunications, Beijing 100876, China (email: \{bohaolu; weizhiqing; dailywu; wlwl; luxi; meidongyang; fengzy\}@bupt.edu.cn).
         Xinrui Zeng is with Beijing University of Aeronautics and Astronautics, Beijing 100191, China (email: \{xr\_zeng\}@buaa.edu.cn)

         Corresponding authors: Zhiqing Wei
}}

\markboth{Journal of \LaTeX\ Class Files,~Vol.~14, No.~8, August~2021}%
{Shell \MakeLowercase{\textit{et al.}}: A Sample Article Using IEEEtran.cls for IEEE Journals}


\maketitle

\begin{abstract}
Utilizing widely distributed communication nodes to achieve environmental reconstruction is one of the significant scenarios for Integrated Sensing and Communication (ISAC) and a crucial technology for 6G. To achieve this crucial functionality, we propose a deep learning based multi-node ISAC 4D environment reconstruction method with Uplink-Downlink (UL-DL) cooperation, which employs virtual aperture technology, Constant False Alarm Rate (CFAR) detection, and Mutiple Signal Classification (MUSIC) algorithm to maximize the sensing capabilities of single sensing nodes. Simultaneously, it introduces a cooperative environmental reconstruction scheme involving multi-node cooperation and Uplink-Downlink (UL-DL) cooperation to overcome the limitations of single-node sensing caused by occlusion and limited viewpoints. Furthermore, the deep learning models Attention Gate Gridding Residual Neural Network (AGGRNN) and Multi-View Sensing Fusion Network (MVSFNet) to enhance the density of sparsely reconstructed point clouds are proposed, aiming to restore as many original environmental details as possible while preserving the spatial structure of the point cloud. Additionally, we propose a multi-level fusion strategy incorporating both data-level and feature-level fusion to fully leverage the advantages of multi-node cooperation. Experimental results demonstrate that the environmental reconstruction performance of this method significantly outperforms other comparative method, enabling high-precision environmental reconstruction using ISAC system.
\end{abstract}

\begin{IEEEkeywords}
Integrated Sensing and Communication, Environmental Reconstruction, Virtual Aperture, Constant False-Alarm Rate, Mutiple Signal Classification, Multi-Node Cooperation, Uplink-Downlink Cooperation, Multi-Level Fusion, Deep Learning.
\end{IEEEkeywords}

\section{Introduction}
\subsection{Background and Motivation}
Integrated Sensing and Communication (ISAC) has been defined as a new feature of next-generation cellular networks by ITU-R as one of the six visions of IMT2030 (6G)\cite{10012421}. The rapid development of 5th Generation (5G), Beyond 5G (B5G) and 6th Generation (6G) wireless communication has conferred communication signals a larger bandwidth and higher frequency bands, which have significantly improved their sensing capabilities. In ISAC scenarios, cellular network devices not only serve as communication nodes to realize the interconnection of everything, but also can be used as sensing nodes to realize functions such as environmental imaging and reconstruction. The multi-node cooperation between the widely distributed communication infrastructure and User Equipment (UE) makes multi-view and omnidirectional environmental imaging and reconstruction possible. Meanwhile, the use of deep neural networks and multilevel fusion strategies can further improve the precision of environmental imaging and reconstruction and maximize the details of the sensing results.

\subsection{Related Works}
Most of the current research on 3D and 4D environmental sensing based on wireless signals utilizes Frequency Modulated Continuous Wave (FMCW) waveform which primarily uses Fast Fourier Transform (FFT)-based algorithms for echo signal processing\cite{sun2020mimo}. Santra {\em et al}.\cite{santra2020ambiguity} and Sun {\em et al}.\cite{sun20214d} proposed the high-resolution Multiple Input Multiple Output (MIMO) radar imaging system based on FMCW waveform. The above works use FFT-based algorithms to process the echoes, which are lacking in terms of environmental sensing resolution compared to MUSIC-based methods. And the FMCW waveform-based environmental sensing scheme is not suitable for ISAC scenarios\cite{niu2024interference, meng2022adaptive}. In recent years the research about single target\cite{chen2023multiple, wang2023coherent, meng2024cramer, 10285442} and multi-target\cite{lu2024integrated} sensing based on Orthogonal Frequency Division Multiplexing (OFDM) signals in ISAC scenarios have achieved good performance. However, there is still a lack of research on OFDM radar-based 4D environmental reconstruction scenarios with massive targets.
Guan {\em et al}.~\cite{guan20213} built a single-node high-resolution 3D radar imaging system using 5G millimeter wave (mmWave) signals. However, due to the existence of occlusions, the system is unable to achieve a complete environmental reconstruction of the sensing area.

Utilizing multi-node cooperation to build a multi-view environmental reconstruction system can overcome the limitations of single-node sensing\cite{wei2024deep, 10226276}. Ren {\em et al}.\cite{ren2021improved} determined the weights of each sensing node by considering factors such as consistency, stability and correlation between parameters. Nuss {\em et al}.\cite{nuss20163d} performed multi-target sensing based on MIMO-OFDM radar through measurement matrix superposition algorithm. Gao {\em et al}.\cite{gao2021reliable} proposed a grid-based probabilistic fusion framework to mitigate the effects of radar ghosting.
The partly multi-radar sensing fusion based approach defines a maximum likelihood function for each position grid, which is estimated as the position corresponding to the maximum value of the function\cite{weiss2011direct,  1288121}. Most of the above related works are cooperative schemes based on multi-base radar, which cannot utilize the inherent advantages of integrating communication node UL with DL and are not applicable to ISAC scenarios as compared to the cooperative schemes based on multiple communication nodes.

Since the sensing capability in ISAC scenario is limited by the frequency and bandwidth resources used by the communication nodes, the beam-pointing situation, and the electromagnetic environment, the point cloud density of the environmental reconstruction results from multi-node cooperation is similarly limited\cite{10256120, yang2024}. In order to make the multi-node cooperative environmental reconstruction result with higher point cloud density to recover the detail information of the original scenario more completely, it is necessary to introduce a point cloud complementation algorithm for low-density reconstruction point cloud density enhancement\cite{10041973, wang2020cascaded}. More recently, several attempts\cite{wang2019dynamic, wang2019deep} have been made to incorporate Graph Convolutional Networks (GCN)\cite{kipf2016semi} to build local graphs in the neighborhood of each point in the point cloud. However, constructing the graph relies on the K-Nearest Neighbor (KNN) algorithm, which is sensitive to the point cloud density\cite{thomas2019kpconv}. In addition to the above, there are almost no open source datasets related to ISAC-based 4D environmental reconstruction, resulting in a lack of work related to low-density point cloud enhancement networks based on ISAC environmental reconstruction scenarios.

\subsection{Contributions}
To solve the above problems, we propose a deep learning based Multi-Node with Downlink-Uplink Cooperative (MNDUC) ISAC 4D environmental reconstruction method. The main contributions and innovations of this paper are summarized as follows.
\begin{itemize}
    \item A Base Station (BS) side ISAC 4D (velocity, 3D position information) environmental reconstruction algorithm based on a virtual receiver array with CFAR detection is proposed. The algorithm uses the standard 5G NR mmWave signal as the signal carrier, processes the echo signal using 2D-FFT with with 2D-MUSIC algorithm, and performs environment scatterers detection based on CFAR.
    \item A multi-node UL-DL cooperative 4D environmental reconstruction method is proposed. This method realizes active sensing at the BS side during the DL time slot and passive sensing at the BS side in cooperation with the UE during the UL time slot. And it achieves data-level ISAC 4D environmental reconstruction result fusion.
    \item A density-enhanced network Attention Gate Gridding Residual Neural Network (AGGRNN) for low-density reconstruction of point clouds for data-level fusion result is proposed. We also apply distributed learning with Multi-modal inputs for feature-level fusion on the foundation of AGGRNN, and build a Multi-View Sensing Fusion Network (MVSFNet) to further optimize the ISAC 4D environmental reconstruction result.
\end{itemize}


We use Chamfer Distance (CD) and F-Score@$1\%$ to measure the performance of the reconstruction results. The simulation results show that, our proposed ISAC 4D environmental reconstruction method has better performance compared to the comparison methods.

The rest of the paper is organized as follows. Section~\ref{sec2} presents our system and signal model. The processing scheme of DL and UL ISAC sensing signal is introduced in Section~\ref{sec3}. In Section~\ref{section4}, the schemes of sensing result fusion and enhancement are proposed. Section~\ref{sec5} discusses our simulation results. Finally, we conclude the work of this paper in Section~\ref{sec6}.

\section{System and Signal Model} \label{sec2}
This section presents the scenario setup, ISAC channel models, and transmit-received signal models for a Multi-Node Downlink-Uplink Cooperative (MNDUC) ISAC imaging method.
\subsection{Scenario Setup of MNDUC ISAC 4D Environmental Reconstruction Method}
\begin{figure}[!htb]
	\centering
	\includegraphics[width=0.8\linewidth]{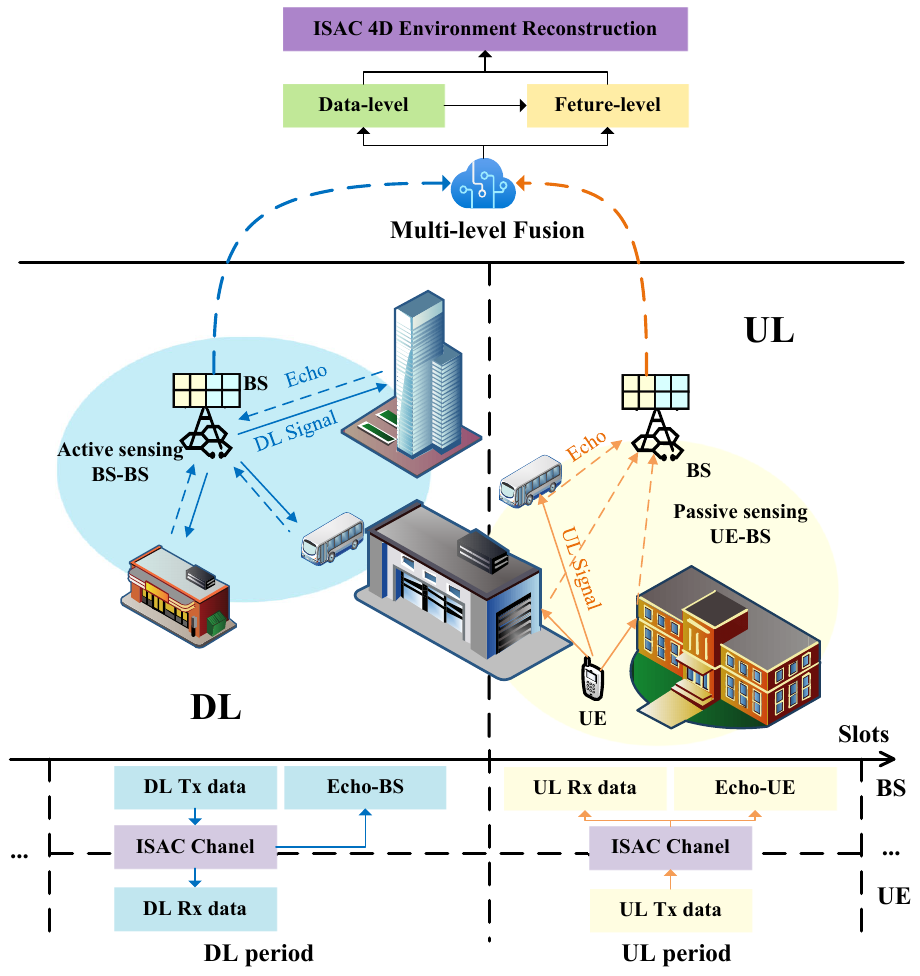}
	\caption{MNDUC ISAC 4D environmental reconstruction scenario.}
	\label{fig:3D_Scenario}
\end{figure}

As shown in Fig.~\ref{fig:3D_Scenario}, we consider a MNDUC ISAC 4D environmental reconstruction scenario between BS, UE, and the environment. The BS is equipped with two spatially well-separated Uniform Planar Arrays (UPAs). The transmit and receive arrays of the BS have dimensions ${{P}_{t}}\times {{Q}_{t}}$ and ${{P}_{r}}\times {{Q}_{r}}$, whose array element spacing and the arrangement will be described in detail below. The UE is equipped with antenna array size ${{P}_{rt}}\times {{Q}_{rt}}$.

The modes of sensing in the DL and UL periods are active and passive sensing, respectively. To obtain a complete environmental reconstruction result, we consider a multi-level fusion strategy to fuse and match the sensing results of multiple sensing nodes in DL and UL periods. 

\subsection{UPAs Model and Virtual Aperture}
\begin{figure}[!htb]
	\centering
	\includegraphics[width=0.8\linewidth]{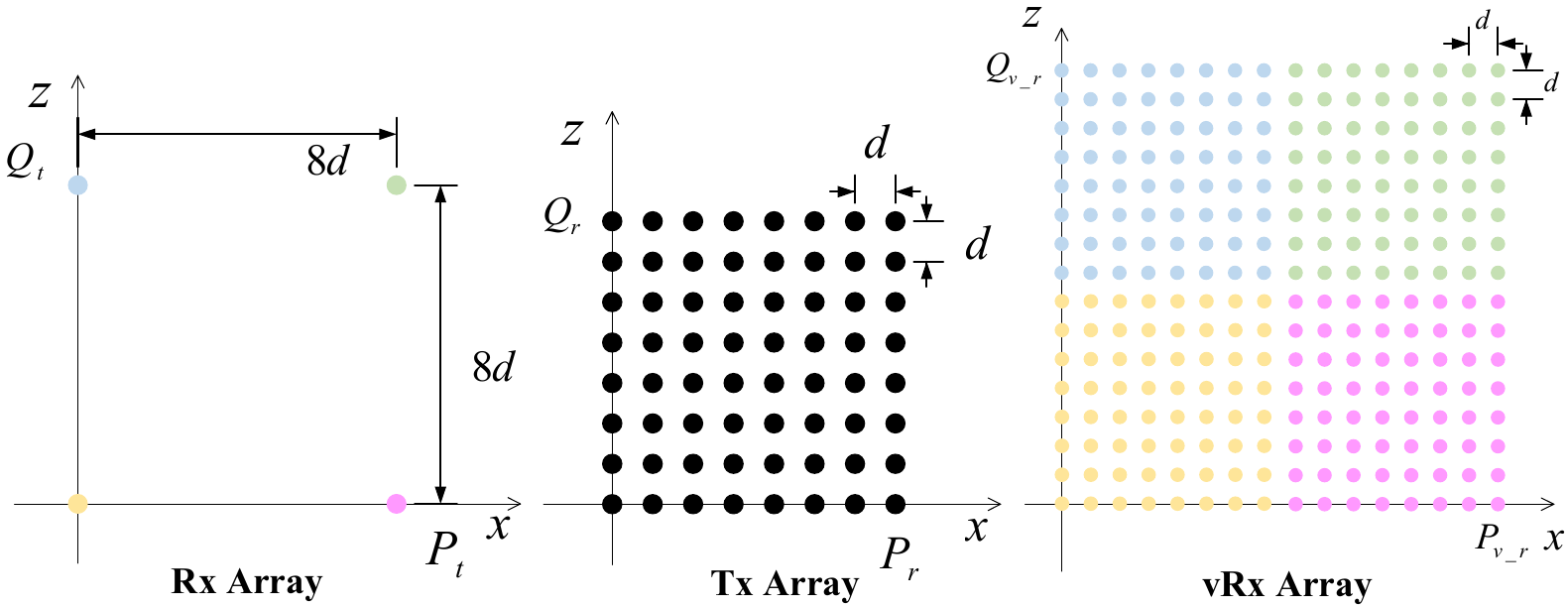}
	\caption{MIMO array design with the transmitting array, the receiving
array and the corresponding virtual Rx (vRx) array.}
	\label{fig:Array}
\end{figure}

We have designed a transmit-receive UPAs as shown in Fig.~\ref{fig:Array} to improve the sensing resolution. The receiving UPA is a ${2\times 2}$ array with an element spacing of ${8d}$, the transmitting UPA is an ${8\times 8}$ array with a spacing of ${d}$, and the corresponding receive UPA is a ${16\times 16}$ array with a spacing of ${d}$.

\begin{figure}[!htb]
	\centering  
	\subfigbottomskip=2pt 
	\subfigcapskip=-5pt 
	\subfigure[Scatterer position.]{
		\includegraphics[width=0.35\linewidth]{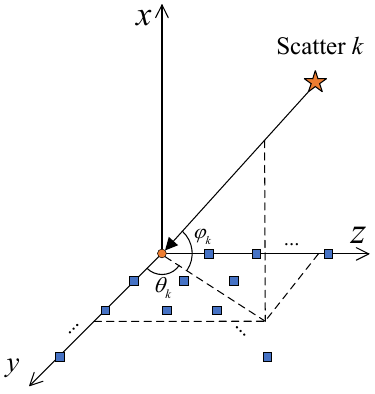}}
        \subfigure[BS position.]{
		\includegraphics[width=0.35\linewidth]{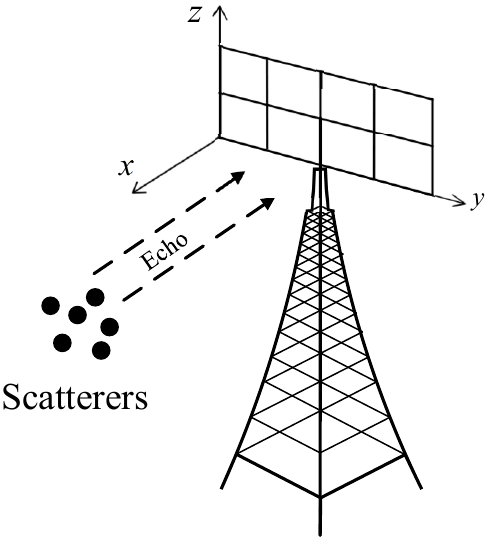}}
	\caption{Schematic diagram of geometric relationships.}
	\label{fig:target_DOA}
\end{figure}
The geometric relationship between the $k$th scattering point and BS is shown in Fig.~\ref{fig:target_DOA}, with the azimuth and pitch angles noted as (${{\theta }_{k}}$, ${{\varphi }_{k}}$). The antenna array element in the virtual receiving array is denoted as ${{A}_{r_v}}({{p}_{r_v}},{{q}_{r_v}})$, and the reference antenna array element is ${{A}_{r_v}}(1,1)$, then the phase difference between ${{A}_{r_v}}({{p}_{r_v}},{{q}_{r_v}})$ and ${{A}_{r_v}}(1,1)$ caused by the $k$th scattering point can be expressed as \cite{10465113}
\begin{equation}
\footnotesize
	\begin{aligned}
		& \Delta {{\phi }_{r_v}}{{\left( {{p}_{r_v}},{{q}_{r_v}} \right)}_{\left( {{\theta }_{k}},{{\varphi }_{k}} \right)}} \\
		&={{e}^{-\xi j2\pi \frac{d\left[ \left( {{p}_{r_v}}-1 \right)\cos {{\theta }_{k}}+\psi \left( {{q}_{r_v}}-1 \right)\sin {{\theta }_{k}} \right]\cos {{\varphi }_{k}}}{\lambda }}} \\ 
		& \left\{ \begin{aligned}
			& \xi =1\ \psi =1,{{\theta }_{k}}<{{90}^{{}^\circ }},{{\varphi }_{k}}<{{90}^{{}^\circ }} \\ 
			& \xi =-1\ \psi =-1,{{\theta }_{k}}>{{90}^{{}^\circ }},{{\varphi }_{k}}<{{90}^{{}^\circ }} \\ 
			& \xi =1\ \psi =-1,{{\theta }_{k}}<{{90}^{{}^\circ }},{{\varphi }_{k}}>{{90}^{{}^\circ }} \\ 
			& \xi =-1\ \psi =1,{{\theta }_{k}}>{{90}^{{}^\circ }},{{\varphi }_{k}}>{{90}^{{}^\circ }} \\ 
		\end{aligned} \right. \\ 
	\end{aligned}	
	\label{eq:position}
\end{equation}

As shown in (\ref{eq:Ak_matrix_DL}), the vRx array phase difference matrix ${{\mathbf{A}}_{k}^{q}}$ in DL sensing period corresponding to the $k$th target is obtained from (\ref{eq:position}).

\begin{equation}
	\centering  
		{{\mathbf{A}}_{k}^{D}}\left( {{p}_{r_v}},{{q}_{r_v}} \right)=\!\Delta {{\phi }_{r_v}}{{\left( {{p}_{r_v}},{{q}_{r_v}} \right)}_{\left( {{\theta }_{k}},{{\varphi }_{k}} \right)}} 
	\label{eq:Ak_matrix_DL}
\end{equation}
As shown in (\ref{eq:Ak_matrix_UL}), the Rx array phase difference matrix ${{\mathbf{A}}_{k}^{q}}$ in UL sensing period corresponding to the $k$th target is obtained from (\ref{eq:position}).
\begin{equation}
	\centering  
		{{\mathbf{A}}_{k}^{U}}\left( {{p}_{r}},{{q}_{r}} \right)=\!\Delta {{\phi }_{r}}{{\left( {{p}_{r}},{{q}_{r}} \right)}_{\left( {{\theta }_{k}},{{\varphi }_{k}} \right)}} 
	\label{eq:Ak_matrix_UL}
\end{equation}
where $q=D$ or $U$ are for DL or UL ISAC signals, respectively.

\subsection{Transmitted ISAC Signals}
To adapt to current mobile wireless communication network systems, OFDM signals are used for both DL and UL signals. The continuous time domain OFDM ISAC signal is defined as\cite{sturm2011waveform}
\begin{equation}
\footnotesize
    {{y}^{q}}\left( t \right)=\sum\limits_{m=0}^{N_{sys}^{q}-1}{\sum\limits_{n=0}^{N_{c}^{q}-1}{s_{\text{Tx}}^{q}\left( n,m \right)}{{e}^{j2\pi \left( {{f}_{c}}+n\Delta {{f}^{q}} \right)t}}\text{rect}\left( \frac{t-mT_{\text{OFDM}}^{q}}{T_{\text{OFDM}}^{q}} \right)}\
\label{eq:OFDM}
\end{equation}
where ${s_{\text{Tx}}^{q}\left( n,m \right)}$ represents the modulated OFDM symbol in the $n$th subcarrier of $m$th OFDM symbol, ${N_{sys}^{q}-1}$ is the number of OFDM symbols and ${N_{c}^{q}-1}$ is the number of subcarriers, ${T_{\text{OFDM}}^{q}}$ is the total duration of the OFDM symbol which satisfies ${T_{\text{OFDM}}^{q}}={{T}^{q}}+{{T}_{\text{CP}}^{q}}$, where $T^{q}$ is the effective OFDM symbol duration and ${T}_{\text{CP}}^{q}$ is the cyclic prefix duration, ${{f}_{c}}$ is the carrier frequency, $\Delta f^{q}={1}/{T^{q}}\;$ is the frequency interval of subcarriers, $\text{rect}\left( \frac{t}{{{T}_{\text{OFDM}}^{q}}} \right)$ is the rectangular function which is equal to 1 for $0\le t\le {{T}_{\text{OFDM}}^{q}}$ and 0 for otherwise.

\subsection{DL and UL ISAC Channel Models}
\begin{figure}[!htb]
	\centering
	\includegraphics[width=0.7\linewidth]{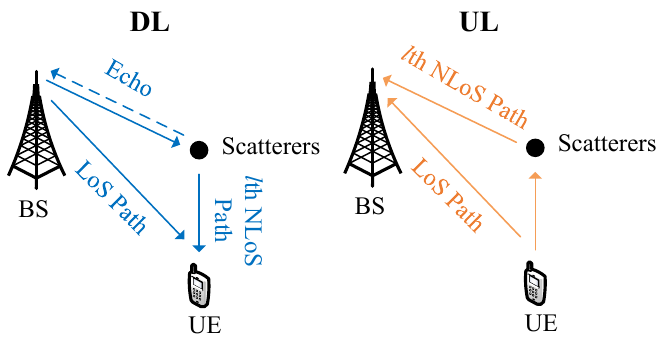}
	\caption{Channel models of DL and UL Channel.}
	\label{fig:channelmodel}
\end{figure}
We denote the UL communication channel as the UL ISAC channel; the DL communication channel is a transposition of the ISAC UL channel due to channel reciprocity; the DL sensing channel consists of the echo path from the scatterers as shown in Fig.~\ref{fig:channelmodel}. We only consider the echo reflected directly from the scatterers.

\subsubsection{UL ISAC Channel Model}
The UL ISAC channel response matrix on the $n$th subcarrier of the $m$th OFDM symbol is defined as\cite{10462908} 
\begin{equation}
    {{\textbf{H}}_{C,n,m}^{U}}=\sum\limits_{l=0}^{L-1}{\left[ \begin{aligned}
      & {{b}_{C,l}}{{e}^{j2\pi \left( {{f}_{c,d,l}} \right)mT_{\text{OFDM}}^{U}{{e}^{-j2\pi n\Delta {{f}^{U}}\left( {{\tau }_{c,l}} \right)}}}} \\ 
     & \times \textbf{a}\left( \textbf{p}_{Rx,l}^{U} \right){{\textbf{a}}^{T}}\left( \textbf{p}_{Tx,l}^{U} \right) \\ 
    \end{aligned} \right]}
\label{eq:ULchannel}
\end{equation}
where $\mathbf{H}_{C,n,m}^{U}\in {{\mathbb{C}}^{{{P}_{t}}{{Q}_{t}}\times {{P}_{rt}}{{Q}_{rt}}}}$, $l=0$ is for the channel response of the LoS (Line of Sight) path, and $l\in \left\{ 1,...,L-1 \right\}$ is for the paths involved with the $l$th scatterer; $\textbf{a}\left( \textbf{p}_{Rx,l}^{U} \right)\in{{\mathbb{C}}^{{{P}_{t}}{{Q}_{t}}\times {1}}}$ and $\textbf{a}\left( \textbf{p}_{Tx,l}^{U} \right)\in{{\mathbb{C}}^{{{P}_{rt}}{{Q}_{rt}}\times {1}}}$ are the steering vectors for UL
reception and transmission, $\textbf{p}_{Rx,l}^{U}$ and $\textbf{p}_{Tx,l}^{U}$ are
the corresponding Angle of Arrival (AoA) and Angle of Departure (AoD), respectively; ${{f}_{c,d,0}}=\frac{{{v}_{0}}}{\lambda }$ and ${{\tau }_{c,0}}=\frac{{{r}_{0,1}}}{c}$ are the Doppler shift and time delay with $v_0$ and ${{r}_{0,1}}$ corresponding to the velocity and range between the UE and BS of the LoS path, respectively; ${{f}_{c,d,l}}={{f}_{d,l,1}}+{{f}_{d,l,2}}$ and ${{\tau }_{c,l}}={{\tau }_{c,l,1}}+{{\tau }_{c,l,2}}$ are the total Doppler shift and time delay of the $l$th NLoS (Non Line of Sight) path, respectively; ${{f}_{d,l,1}}=\frac{{{v}_{r,l,1}}}{\lambda }$ and ${{f}_{d,l,2}}=\frac{{{v}_{r,l,2}}}{\lambda }$ are the Doppler shifts between the UE and
the $l$th scatterer, and between the $l$th scatterer and the BS,
respectively, with ${{{v}_{r,l,1}}}$ and ${{{v}_{r,l,2}}}$ being the velocities; ${{\tau }_{c,l,1}}=\frac{{{r}_{l,1}}}{c}$ and ${{\tau }_{c,l,2}}=\frac{{{r}_{l,2}}}{c}$ are the time
delays between the user and the $l$th scatterer, and between
BS and the $l$th scatterer, respectively, with ${{{r}_{l,1}}}$ and ${{{r}_{l,2}}}$ being
the ranges. The attenuation  ${{b}_{C,l}}=\sqrt{\frac{{{\lambda }^{2}}}{{{\left( 4\pi {{r}_{0,1}} \right)}^{2}}}},l=0$ for the LoS path and ${{b}_{C,l}}={{\beta }_{C,l}}\times \sqrt{\frac{{{\lambda }^{2}}}{{{\left( 4\pi  \right)}^{3}}{{r}_{l,1}}^{2}{{r}_{l,2}}^{2}}},l\in \left\{ 1,...,L-1 \right\}$ for the NLoS paths, where ${{\beta }_{C,l}}$ is the reflecting factor of the $l$th scatterer, following $\mathcal{C}\mathcal{N}\left( 0, \sigma _{C\beta ,l}^{2} \right)$.

\subsubsection{DL Communication Channel Model}
From the channel reciprocity, the transpose of the UL communication channel response is the DL communication channel response.
\begin{equation}
    {\textbf{H}}_{C,n,m}^{D}=\sum\limits_{l=0}^{L-1}{\left[ \begin{aligned}
      & {{b}_{C,l}}{{e}^{j2\pi \left( {{f}_{c,d,l}} \right)mT_{\text{OFDM}}^{D}{{e}^{-j2\pi n\Delta {{f}^{D}}\left( {{\tau }_{c,l}} \right)}}}} \\ 
     & \times \textbf{a}\left( \textbf{p}_{Rx,l}^{D} \right){{\textbf{a}}^{T}}\left( \textbf{p}_{Tx,l}^{D} \right) \\ 
    \end{aligned} \right]}
\label{eq:DLchannelcom}
\end{equation}
where $\mathbf{H}_{C,n,m}^{D}\in {{\mathbb{C}}^{{{P}_{t}}{{Q}_{t}}\times {{P}_{rt}}{{Q}_{rt}}}}$, $\textbf{a}\left( \textbf{p}_{Rx,l}^{D} \right)\in{{\mathbb{C}}^{{{P}_{t}}{{Q}_{t}}\times {1}}}$ and $\textbf{a}\left( \textbf{p}_{Tx,l}^{D} \right)\in{{\mathbb{C}}^{{{P}_{rt}}{{Q}_{rt}}\times {1}}}$ are the steering vectors for DL transmission, $\textbf{p}_{Rx,l}^{D}=\textbf{p}_{Tx,l}^{U}$ and $\textbf{p}_{Tx,l}^{D}=\textbf{p}_{Rx,l}^{U}$ are the DL communication AoA and AoD, respectively.

\subsubsection{DL Sensing Channel Model}
The DL sensing channel response matrix on the $n$th subcarrier of the $m$th OFDM symbol is defined as\cite{10462908}
\begin{equation}
    {\textbf{H}}_{S,n,m}^{D}=\sum\limits_{l=0}^{L-1}{\left[ \begin{aligned}
      & {{b}_{S,l}}{{e}^{j2\pi \left( {{f}_{s,l,1}} \right)mT_{\text{OFDM}}^{D}{{e}^{-j2\pi n\Delta {{f}^{D}}\left( {{\tau }_{s,l}} \right)}}}} \\ 
     & \times \textbf{a}\left( \textbf{p}_{Rx,l}^{DS} \right){{\textbf{a}}^{T}}\left( \textbf{p}_{Tx,l}^{D} \right) \\ 
    \end{aligned} \right]}
\label{eq:DLchannelsensing}
\end{equation}
where $\textbf{a}\left( \textbf{p}_{Rx,l}^{DS} \right) \in {{\mathbb{C}}^{{{P}_{r_v}}{{Q}_{r_v}}\times 1}}$ and ${{\textbf{a}}^{T}}\left( \textbf{p}_{Tx,l}^{D} \right) \in {{\mathbb{C}}^{{{P}_{r_v}}{{Q}_{r_v}}\times 1}}$ are the steering vectors for DL echo
reception and DL transmission, respectively; $\textbf{p}_{Rx,l}^{DS}$ and $\textbf{p}_{Tx,l}^{D}$ are the AoD and AoA of the ISAC sensing receiver and transmitter array, respectively. Since mmWave arrays are typically small and equivalent to virtual receiver arrays during the DL sensing period as shown in Fig.~\ref{fig:Array}, $\textbf{p}_{Rx,l}^{DS}=\textbf{p}_{Tx,l}^{D}$. 
Where ${f}_{s,0,1}=\frac{2{{v}_{0}}}{\lambda }$ and ${f}_{s,l,1}=\frac{2{{v}_{r,l,2}}}{\lambda }$ are the total Doppler shift of the $l$th echo path with $v_0$ and ${v}_{r,l,2}$ being the velocities; ${\tau }_{s,0}=\frac{2{{r}_{0,1}}}{c}$ and ${\tau }_{s,l}=\frac{2{{r}_{l,2}}}{c}$ are the time delays of the $l$th echo path with ${r}_{0,1}$ and ${r}_{l,2}$ being the ranges. ${{b}_{S,l}}={{\beta }_{S,l}}\times \sqrt{\frac{{{\lambda }^{2}}}{{{\left( 4\pi  \right)}^{3}}{{d}_{l,2}}^{4}}},l\in \left\{ 0,...,L-1 \right\}$ is the attenuation for the echo path,where ${{\beta }_{S,l}}$ is the reflecting factor of the $l$th scatterer, following $\mathcal{C}\mathcal{N}\left( 0, \sigma _{S\beta ,l}^{2} \right)$.

\subsection{Received ISAC Signals}
\subsubsection{Received Communication Signals}
The received communication signal on the $n$th subcarrier of the $m$th OFDM symbol is defined as\cite{10465113, 10462908}
\begin{equation}
\footnotesize
    \begin{aligned}
      & \textbf{y}_{C,n,m}^{q}=s_{Tx}^{q}\left( n,m \right)\textbf{H}_{C,n,m}^{q}\text{w}_{Tx}^{q}+\textbf{n}_{t,n,m}^{q} \\ 
     & =s_{Tx}^{q}\left( n,m \right)\sum\limits_{l=0}^{L-1}{\left[ \begin{aligned}
      & {{b}_{C,l}}{{e}^{j2\pi \left( {{f}_{c,d,l}} \right)mT_{\text{OFDM}}^{q}{{e}^{-j2\pi n\Delta {{f}^{q}}\left( {{\tau }_{c,l}} \right)}}}} \\ 
     & \times \textbf{a}\left( \textbf{p}_{Rx,l}^{q} \right)\chi _{Tx,l}^{q} \\ 
    \end{aligned} \right]}+\textbf{n}_{t,n,m}^{q} \\ 
    \end{aligned}
\label{eq:received_com}
\end{equation}
where $q=D$ or $U$ are for DL or UL ISAC signals, respectively; $\textbf{n}_{t,n,m}^{q}$ is the noise vector and each element of $\textbf{n}_{t,n,m}^{q}$ follow $\mathcal{C}\mathcal{N}\left( 0, \sigma _{N}^{2} \right)$; $\textbf{y}_{C,n,m}^{q}, \textbf{n}_{t,n,m}^{q} \in {\mathbb{C}}^{{{P}_{t}}{{Q}_{t}}\times {1}}$ and $\textbf{y}_{C,n,m}^{q}, \textbf{n}_{t,n,m}^{q} \in {\mathbb{C}}^{{{P}_{r\_t}}{{Q}_{r\_t}}\times {1}}$ when $q=U$ and $q=D$, respectively; ${s_{\text{Tx}}^{q}\left( n,m \right)}$ represents the modulated OFDM symbol; $\text{w}_{Tx}^{q}$ is the transmit Beamforming (BF) vector, and $\chi _{Tx,l}^{q}={{\textbf{a}}^{T}}\left( \textbf{p}_{Tx,l}^{q} \right)\text{w}_{Tx}^{q}$ is the transmit BF gain.
\subsubsection{Received Sensing Signals}
The received signal when the OFDM signal shown in (\ref{eq:OFDM}) reflected by the $k$th scatterer is defined as\cite{10465113}
\begin{equation}
    \begin{aligned}
        & y_{k}^{q}\left( t \right)={{G}_{k}}\sum\limits_{m=0}^{{{N}_{sym}^{q}}-1}{{{e}^{j2\pi {{{f}_{d}}\left( k \right)}t}}\sum\limits_{n=0}^{{{N}_{c}^{q}}-1}{{{s}_{Rx}^{q}}\left( n,m,k \right)}}\times  \\ 
        & {{e}^{j2\pi \left( {{f}_{c}}+n\Delta {{f}^{q}} \right)\left( t-\frac{2{{R}_{k}}}{c} \right)}}\times \text{rect}\left( \frac{t-m{{T}_{\text{OFDM}}^{q}}-\frac{2{{R}_{k}}}{c}}{{{T}_{\text{OFDM}}^{q}}} \right)  
    \end{aligned}
\label{eq:ofdm_reflected}
\end{equation}
where $R_k$ and ${{f}_{d}}\left( k \right)$ is the range and Doppler shift of $k$th scatter, respectively. $G_k$ represents the attenuation factor associated with the path loss, radar cross section (RCS) of the $k$th scatter.
The relationship between the received modulation symbols ${{{s}_{Rx}^{q}}\left( n,m,k \right)}$ and the transmitted modulation symbol ${{{s}_{Tx}^{}q}\left( n,m\right)}$ can be defined as
\begin{equation}
    {{s}_{Rx}^{q}}\left( n,m,k \right)={{G}_{k}}{{s}_{Tx}^{q}}\left( n,m \right){{e}^{-j2\pi {{f}_{n}^{q}}\frac{2{{R}_{k}}}{c}}}{{e}^{j2\pi {{{f}_{d}}\left( k \right)mT_{\text{OFDM}}^{q}}}}
    \label{eq:Rx_Tx_reletionship}
\end{equation}
where ${f}_{n}^{q}={{f}_{c}}+n\Delta {{f}^{q}}$.

The CSI(Channel State Information) matrix information matrix ${{\textbf{s}}_{g}^{q}}$ is defined as
\begin{equation}
    {{\textbf{s}}_{g}^{q}}=\frac{{{\textbf{s}}_{Rx}^{q}}} {{{\textbf{s}}_{Tx}^{q}}}={{G}_{k}}\left( {\textbf{k}}_{r}^{q}\otimes {\textbf{k}}_{d}^{q} \right)
    \label{eq:matrix_info_rdm}
\end{equation}
where ${\textbf{k}}_{r}^{q}=\left( 0,{{e}^{-j2\pi \Delta f\frac{2{{R}_{k}}}{c}}},\cdots ,{{e}^{-j2\pi \left( {{N}_{c}^{q}}-1 \right)\Delta f\frac{2{{R}_{k}}}{c}}} \right)$ and ${\textbf{k}}_{d}^{q}=\left( 0,{{e}^{j2\pi {{T}_{\text{OFDM}}^{q}}{{f}_{d}}\left( k \right)}},\cdots ,{{e}^{j2\pi ({{N}_{sym}^{q}}-1){{T}_{\text{OFDM}}^{q}}{{f}_{d}}\left( k \right)}} \right)$ are the two vectors carrying the Doppler and range information. $\otimes$ refers to a dyadic product\cite{sturm2011waveform}.

The Direction of Arrival (DoA) information matrix ${\mathbf{A}}_{S}^{q}$ of all array elements is defined as
\begin{equation}
    {\mathbf{A}}_{S}^{q}=\sum\limits_{k}{{{\textbf{s}}_{g}^{q}}{{\mathbf{A}}_{k}^{q}}} 
    \label{eq:Array_signal_sumk}
\end{equation}
It is worth noting that during the DL sensing period, the DoA satisfies the Far-field assumption that the AoAs are in the same direction and thus the vRx arrays can be used, while in the UL sensing period, the above assumption is not satisfied and thus only real aperture Rx arrays could be used\cite{balanis2016antenna}. 

The $\textbf{A}_{S}^{q}$ in our simulation work exists in the form of 4D arrays which contain the dimensions of fast time dimension, slow time dimension, horizontal antenna array dimension and vertical antenna array dimension. In actual ISAC 4D environmental reconstruction system, $\textbf{A}_{S}^{q}$ represents the environmental scatterer reflection signal received by UPA of BS.

\section{Methodology of ISAC Sensing Signal Processing} \label{sec3}
\begin{figure}[!htb]
	\centering
	\includegraphics[width=0.8\linewidth]{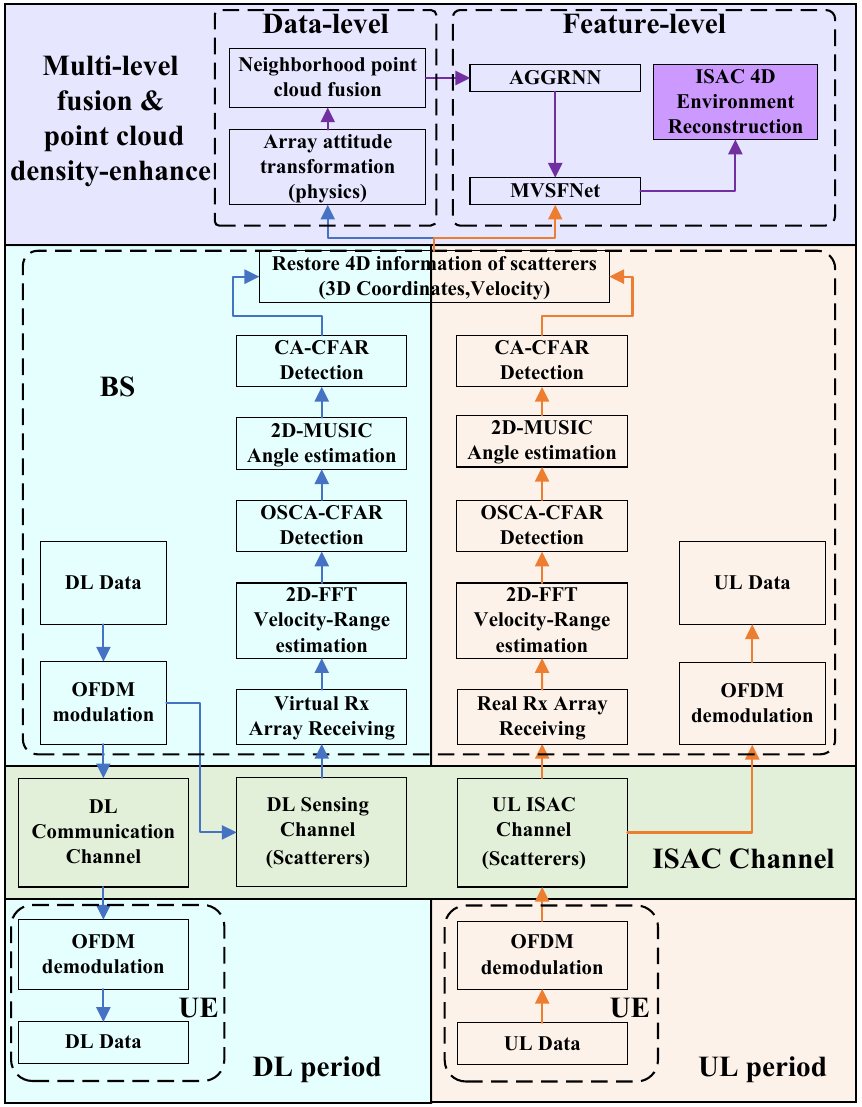}
	\caption{The illustration of DL and UL ISAC signal processing and multi-level fusion strategy.}
	\label{fig:sysflow}
\end{figure}
\begin{figure}[!htb]
	\centering
	\includegraphics[width=0.65\linewidth]{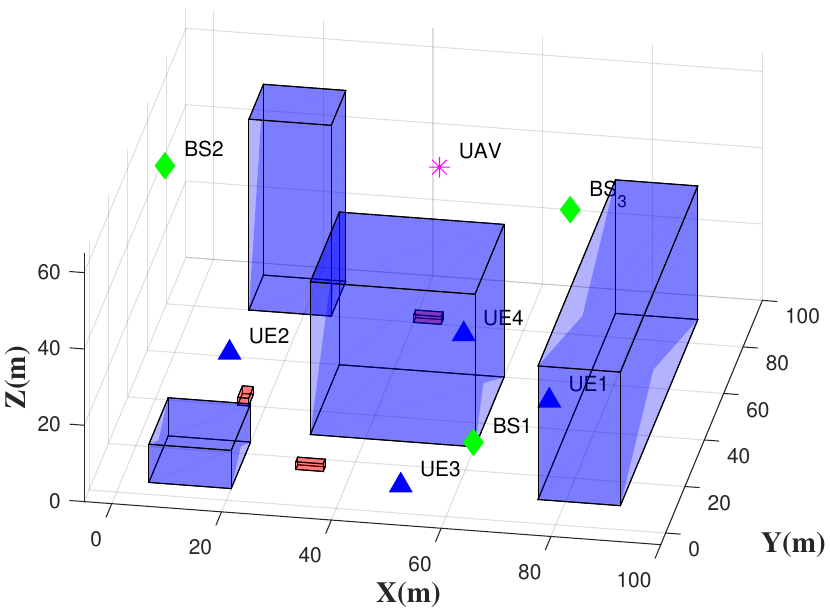}
	\caption{Simulation scenario modeling.}
	\label{fig:abstract3d}
\end{figure}

The DL and UL ISAC sensing signal processing scheme is shown in Fig.~\ref{fig:sysflow}. The multi-level fusion strategy and enhancement method of the sensing results will be shown in Section~\ref{section4}. In this section, we first introduce the DL sensing signal processing algorithm, then we introduce the UL sensing signal processing algorithm. We abstract the description of the scenario in Fig.~\ref{fig:3D_Scenario} as shown in Fig.~\ref{fig:abstract3d}.

\subsection{DL ISAC Sensing Signal Processing} \label{DLsignalproc}
In this section, we first introduce the Doppler and range estimation based on 2D-FFT and OSCA (Ordered Statistics \& Cell Averaging)-CFAR algorithms, then introduce the DoA estimation based on 2D-MUSIC and CA (Cell Averaging)-CFAR algorithms.
\subsubsection{Range and Doppler Estimation}
${\textbf{k}}_{r}^{q}, q=D$ is the linear phase shift vector along the fast time axis caused by range $R$. We initially process the received signal of each vRX antenna array element by using (\ref{eq:matrix_info_rdm}) and then performing the Inverse Discrete Fourier Transform (IDFT) along the subcarrier dimension\cite{sturm2011waveform}.
\begin{equation}
	\begin{aligned}
		& r^{D}\left( \alpha  \right)=\text{IDFT}\left[ {{k}_{r}^{D}}\left( n \right) \right]=\frac{1}{{{N}_{c}^{D}}}\sum\limits_{n=0}^{{{N}_{c}^{D}}-1}{{{k}_{r}^{D}}\left( n \right){{e}^{jn\alpha \frac{2\pi }{{{N}_{c}^{D}}}}}} \\ 
		& =\frac{1}{{{N}_{c}^{D}}}\sum\limits_{n=0}^{{{N}_{c}^{D}}-1}{{{e}^{-j2\pi n\Delta f^{D}\frac{2R}{c}}}{{e}^{jn\alpha \frac{2\pi }{{{N}_{c}^{D}}}}}},\alpha =0,\ldots ,{{N}_{c}^{D}}-1  
	\end{aligned}
	\label{eq:kr_fft}
\end{equation}
where $ r^{D}\left( \alpha  \right)$ obtains peak value when $\alpha =\left\lfloor \frac{2R\Delta f^{D}{{N}_{c}^{D}}}{c} \right\rfloor $. 

${\textbf{k}}_{d}^{q}, q=D$ is the linear phase shift vector along the slow time axis caused by velocity $v_k$. Similarly $v_k$ can be estimated by performing a Discrete Fourier Transform (DFT) along the OFDM symbol dimension\cite{sturm2011waveform}.
\begin{equation}
	\begin{aligned}
		 &  v^{D}\left( \beta  \right)=\text{DFT} \left[ {{k}_{d}^{D}}\left( m \right) \right]=\sum\limits_{m=0}^{{{N}_{sym}^{D}}-1}{{{k}_{d}^{D}}\left( m \right){{e}^{-jm\beta \frac{2\pi }{{{N}_{sym}^{D}}}}}} \\ 
		&  = \sum\limits_{m=0}^{{{N}_{sym}^{D}}-1} {{{e}^{j2\pi m{{T}^{D}_{\text{OFDM}}}\frac{2{{v}_{k}}{{f}_{c}^{D}}}{c}}}{{e}^{-jm\beta \frac{2\pi }{{{N}_{sym}^{D}}}}}},\beta =0,\ldots ,{{N}_{sym}^{D}}-1 \\ 
	\end{aligned}
	\label{eq:kd_fft}
\end{equation}
where ${v^{D}\left( \beta  \right)}$ obtains peak value when $\beta$ in (\ref{eq:kd_fft}) satisfies $\beta =\left\lfloor \frac{2{{v}_{k}}{{f}_{c}^{D}}{{T}^{D}_{\text{OFDM}}}{{N}_{sym}^{D}}}{c} \right\rfloor $. 

\begin{figure}[!htb]
	\centering  
	\subfigure[2D OSCA-CFAR.]{
		\includegraphics[width=0.31\linewidth]{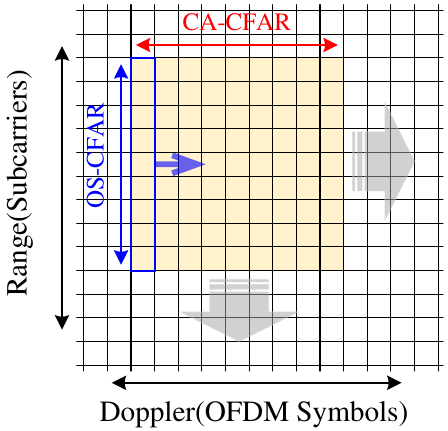}}
 	\subfigure[RDM.]{
		\includegraphics[width=0.31\linewidth]{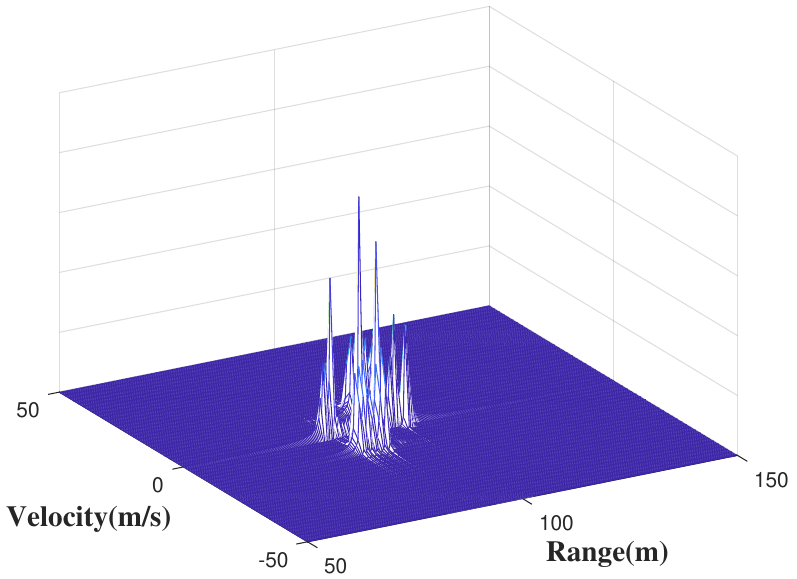}}
	\subfigure[Threshold.]{
		\includegraphics[width=0.31\linewidth]{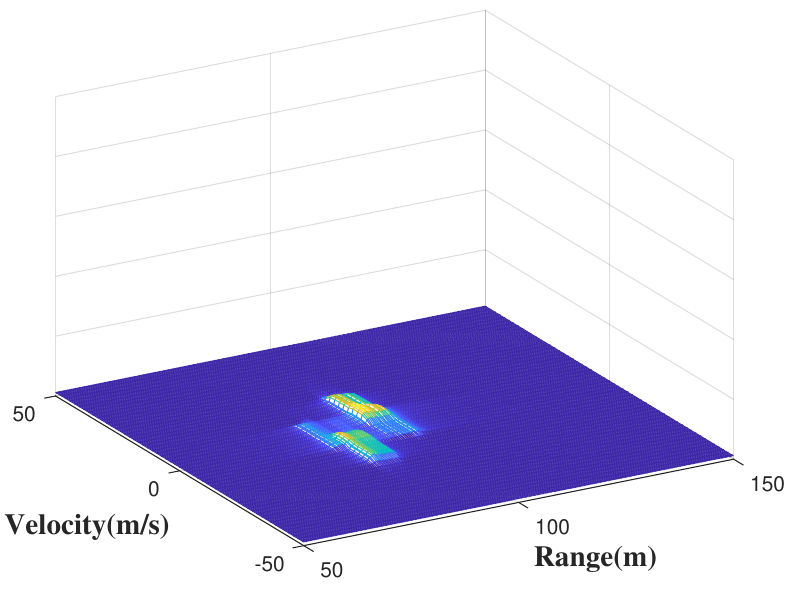}}
	\caption{(a) is the sliding reference window design for 2D OSCA-CFAR, (b) and (c) are the Radar Doppler Map (RDM) and its detection threshold, respectively.}
	\label{fig:RDMproc}
\end{figure}

The RDM can be obtained by processing $\textbf{s}_g^{D}$ in (\ref{eq:matrix_info_rdm}) using the 2D-FFT algorithm, denoted as $\textbf{s}_{_{g}}^{D,R,v}$. As shown in Fig.~\ref{fig:RDMproc} (a), we use the 2D OSCA-CFAR detection algorithm to estimate the range and velocity information of the scatterers by detecting the RDM. We select a reference window of size $9\times9$ within the RDM to show the detection process.
\begin{itemize}
\item We apply 1D OS-CFAR to the sliding reference window by column and the $\gamma$th value of each colum is selected as the estimated noise intensity of this sliding window. We take $\gamma=\left\lfloor 0.75N \right\rfloor $, where $N$ is the number of reference cells durning the sliding reference window.
\item As shown in (\ref{eq:OSCA_CFAR}), the estimated values of noise intensity ${{\bar{\mu }}_{\left( \gamma \right)}}$ are averaged by row as the noise threshold of the Center Detection Unit (CUT) in reference window.
\begin{equation}
	{{\bar{\mu }}_{\left( \gamma \right)}}=\frac{1}{N}\sum\limits_{n=1}^{N}{{{X}_{\left( \gamma \right),n}}}
	\label{eq:OSCA_CFAR}
\end{equation}
where ${X}_{\left( \gamma \right),n}$ represents the noise reference value selected by OS-CFAR, since the size of the sliding reference window we select is $9\times9$, $N=9$.
\item The detection threshold factor ${{T}_{f}}$ of the selected CUT is only related to the number of reference units $N$ with the specified false alarm probability ${{P}_{fa}}$, which is given by  (\ref{eq:threshold_factor}). 
\begin{equation}
	{{T}_{f}}={{\left( {{P}_{fa}} \right)}^{-\frac{1}{N}}}-1
	\label{eq:threshold_factor}
\end{equation}
We can calculate the detection threshold for the selected CUT $T$ by $T={{T}_{f}}\times {{\bar{\mu }}_{\left( \gamma \right)}}$, and the 2D adaptive detection threshold can be obtained by applying the above operations to all CUTs in the RDM as shown in Fig.~\ref{fig:RDMproc}.
\end{itemize}

\subsubsection{DoAs Estimation}
\begin{figure}[!htb]
	\centering  
	\subfigure[2D CA-CFAR.]{
		\includegraphics[width=0.31\linewidth]{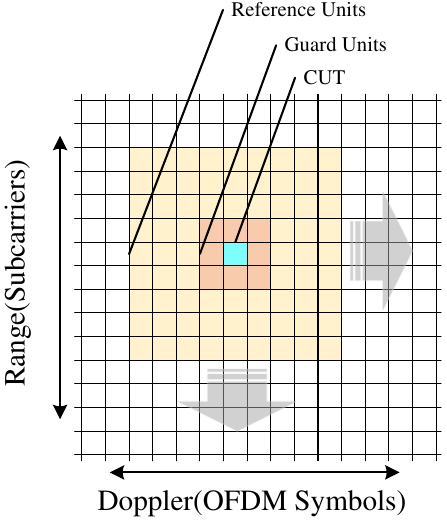}}
 	\subfigure[Pseudo-spectrum.]{
		\includegraphics[width=0.31\linewidth]{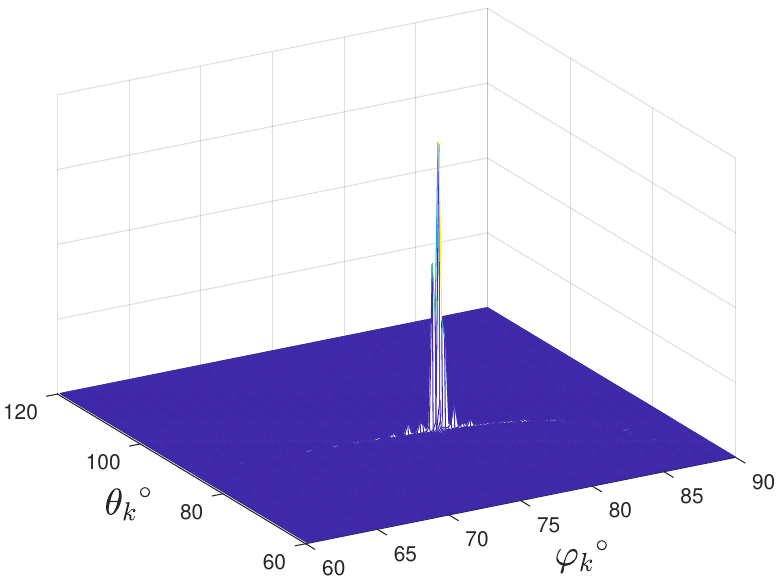}}
	\subfigure[Threshold.]{
		\includegraphics[width=0.31\linewidth]{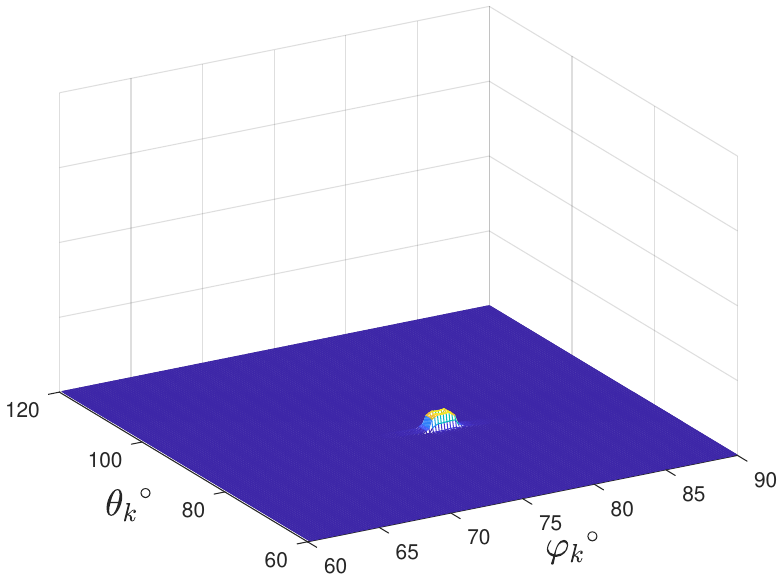}}
	\caption{(a) is the sliding reference window design for 2D CA-CFAR, (b) and (c) are the MUSIC pseudo-spectrum and its detection threshold, respectively.}
	\label{fig:MUSICproc}
\end{figure}
The peak of the RDM indicates the presence of targets with velocity $v$ and range $R$ but $\varphi, \theta$ of the DoAs are unknown.The linear phase shift of the corresponding $\textbf{s}_{_{g}}^{D,R,v}$ value due to $R$ and $v$ has been removed\cite{sturm2011waveform}, which carries only the linear phase shift caused by the wave range difference of the receiving antenna array elements. As shown in (\ref{eq:A_km}), we combine the RDM peaks detected by 2D OSCA-CFAR of all receiving antenna elements into ${{k}_{m}}\left( {{k}_{m}}<k \right)$ manifolds for MUSIC-based DoA estimation.
\begin{equation}
    \footnotesize
		{\mathbf{A}_{{{k}_{m}}}^{D}}\left( {{p}_{r_v}},{{q}_{r_v}} \right)=\sum\limits_{t}{{{G}_{t}}{\mathbf{A}_{t}^{D}}} =\sum\limits_{t}{{{G}_{t}}\Delta {{\phi }_{r_v}}{{\left( {{p}_{r_v}},{{q}_{r_v}} \right)}_{\left( {{\theta }_{t}},{{\varphi }_{t}} \right)}}} 
	\label{eq:A_km}
\end{equation}

\begin{figure}[!htb]
	\centering
	\includegraphics[width=0.7\linewidth]{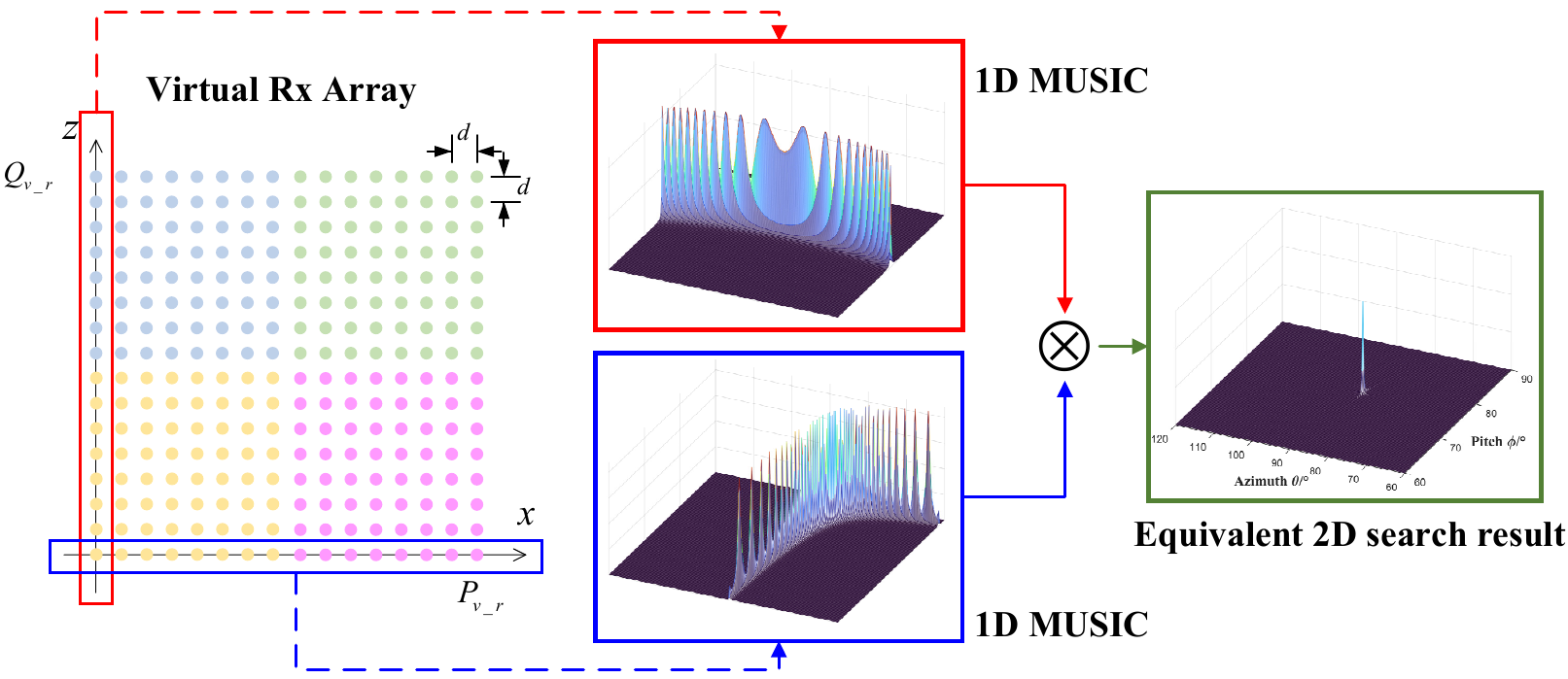}
	\caption{Using the pseudo-spectrum multiplication of two 1D-MUSIC searches instead of 2D-MUSIC searches to reduce complexity.}
	\label{fig:music1d2d}
\end{figure}

We take ${{\left( {\mathbf{A}_{{{k}_{m}}}^{D}} \right)}_{1,:}}$ and ${{\left( {\mathbf{A}_{{{k}_{m}}}^{D}} \right)}_{:,1}}$ of ${\mathbf{A}_{{{k}_{m}}}^{D}}$ to construct the searching manifolds, which are related to the 2D angle ${\mathbf{p}}=\left( \theta ,\varphi  \right)$ of the targets simultaneously, and eventually just multiply the two search results to get the 2D angle information of $t$ targets in ${\mathbf{A}_{{{k}_{m}}}^{D}}$. The visualized image of the MUSIC search results of the above two sets of 1D line arrays and their multiplication is shown in Fig.~\ref{fig:music1d2d}.

The algorithmic flow of the MUSIC DoA search through ${{\left( {\mathbf{A}_{{{k}_{m}}}^{D}} \right)}_{:,1}}$ is as follows, and the processing of ${{\left( {\mathbf{A}_{{{k}_{m}}}^{D}} \right)}_{1,:}}$ can be obtained in the same way.
\begin{itemize}
\item We need to ensure the availability of the MUSIC algorithm by spatial smoothing algorithm to decoherence the individual scatterer echoes. As shown in Fig.~\ref{fig:foward_smoothing}, we can define the forward spatial smoothing matrix ${{\mathbf{R}}_{f}}=\frac{1}{L}\sum\limits_{l=1}^{L}{{\mathbf{R}}_{l}^{f}}$, and similarly the backward spatial smoothing matrix ${{\mathbf{R}}_{b}}=\frac{1}{L}\sum\limits_{l=1}^{L}{{\mathbf{R}}_{l}^{b}}$, where ${{\mathbf{R}}_{l}^{f}}$ and ${{\mathbf{R}}_{l}^{b}}$ are the covariance matrices of the subarray and $L$ is the number of subarray elements. Then use the average of them ${\mathbf{R_X}}=\frac{{{\mathbf{R}}_{f}}+{{\mathbf{R}}_{b}}}{2}$ to replace the original covariance matrix.

\begin{figure}[!htb]
	\centering
	\includegraphics[width=0.35\linewidth]{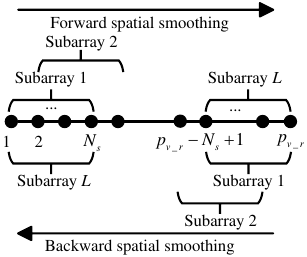}
	\caption{Schematic of the forward and backward space smoothing algorithm.}
	\label{fig:foward_smoothing}
\end{figure}

\item We can get (\ref{eq:eig}) by applying eigenvalue decomposition to ${\mathbf{R_X}}$.
\begin{equation}
	\left[ {{\mathbf{U}}_{x}},{{\mathbf{\Sigma }}_{x}} \right]=\text{eig}\left( {{\mathbf{{R}_{X}}}} \right)
	\label{eq:eig}
\end{equation}
where ${{\mathbf{\Sigma }}_{x}}$ is the real-valued eigenvalue diagonal matrix and ${{\mathbf{U}}_{x}}$ is the orthogonal eigenmatrix. We obtain the number of targets $k_s$ by performing 1D CA-CFAR detection on ${{\mathbf{\Sigma }}_{x}}$.

\item Construct ${{U}_{N}}={{U}_{x}}\left( :,1:{{N}_{s}}-{{N}_{x}} \right)$ as the noise subspace basis.Then we can obtain the spatial angular spectrum function as \cite{haardt2014subspace}
\begin{equation}
	{{f}_{a}}\left( {\mathbf{p}};{{\mathbf{U}}_{N}} \right)={{\mathbf{a }}^{\text{H}}}\left( {\mathbf{p}} \right){{\mathbf{U}}_{N}}{{\left( {{\mathbf{U}}_{N}} \right)}^{\text{H}}}{\mathbf{a}} \left( {\mathbf{p}} \right)
	\label{eq:spatial_spectrum}
\end{equation}
where  ${\mathbf{p}}=\left( \theta ,\varphi  \right)$ is the 2D angle, ${\mathbf{a}}\left( {\mathbf{p}} \right)$ is given in ${{\mathbf{A}}_{k}^{D}}$ as shown in (\ref{eq:Ak_matrix_DL}). The spatial pseudo-spectrum is represented as \cite{haardt2014subspace}
\begin{equation}
	{\textbf{S}_{a}^{row}}\left( {\mathbf{p}};{{\mathbf{U}}_{N}} \right)={{\left[ {{\mathbf{a}}^{\text{H}}}\left( {\mathbf{p}} \right){{\mathbf{U}}_{N}}{{\left( {{\mathbf{U}}_{N}} \right)}^{\text{H}}}{\mathbf{a}} \left( {\mathbf{p}} \right) \right]}^{-1}}
	\label{eq:music_spectrum}
\end{equation}
Similarly we can obtain ${{\textbf{S}_{a}^{col}}\left( {\mathbf{p}};{{\mathbf{U}}_{N}} \right)}$ by taking the above operation for ${{\left( {\mathbf{A}_{{{k}_{m}}}^{D}} \right)}_{1,:}}$. 
\end{itemize}
We obtained the estimated DoAs by performing 2D CA-CFAR detection on ${{\textbf{S}_{a}}\left( {\mathbf{p}};{{\mathbf{U}}_{N}} \right)}={{\textbf{S}_{a}^{col}}\left( {\mathbf{p}};{{\mathbf{U}}_{N}} \right)}\odot {{\textbf{S}_{a}^{row}}\left( {\mathbf{p}};{{\mathbf{U}}_{N}} \right)}$, $\odot$ represents Hadamard product and ${{\textbf{S}_{a}}\left( {\mathbf{p}};{{\mathbf{U}}_{N}} \right)}$ is shown in Fig.~\ref{fig:music1d2d}. 

As shown in Fig.~\ref{fig:MUSICproc}(a), the noise of the 2D CA-CFAR is determined by averaging the $N_R$ reference cells excluding the protection cell and the CUT.
\begin{equation}
	\bar{\mu }=\frac{1}{{{N}_{R}}}\sum\limits_{n=1}^{{{N}_{R}}}{{{X}_{n}}}
	\label{eq:CA-CFAR}
\end{equation}
where ${{X}_{n}}$ represents the reference unit value. The detection threshold of this CUT can be calculated from $T={{T}_{f}}\times {{\bar{\mu }}}$ where ${{T}_{f}}$ is given by (\ref{eq:threshold_factor}), and the 2D adaptive detection threshold can be obtained by applying the above operations to all CUTs in ${{\textbf{S}_{a}}\left( {\mathbf{p}};{{\mathbf{U}}_{N}} \right)}$ as shown in Fig.~\ref{fig:MUSICproc} (b)-(c).

\subsection{UL ISAC Sensing Signal Processing} \label{ULproc}
\begin{figure}[!htb]
	\centering
	\includegraphics[width=0.7\linewidth]{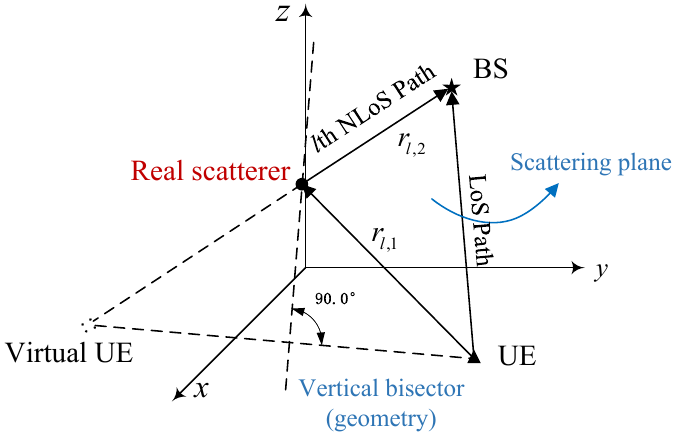}
	\caption{The geometric relationship between the UE, BS, and scatterer.}
	\label{fig:ULsensing}
\end{figure}
The positional relationship of the scatterers and the UE Tx array does not satisfy the far-field assumption\cite{balanis2016antenna}. Therefore we can't receive the echoes of UL signal with the vRx array during the UL sensing period, but only with the BS-side real Rx array. 

The UL sensing based on UL signal echoes from $l$ NLos paths. The methodology for estimating the scatterers' 4D information is essentially the same as shown in \ref{DLsignalproc}. As shown in Fig.~\ref{fig:ULsensing}, the DoAs of the UL signal echoes can be correctly estimated, but due to the time delay $\tau $ of the $l$th path sourced with the two distances $r_{l,1}$ and $r_{l,2}$, the scatterer's range information from the BS cannot be correctly estimated. The position of the scatterer of the $l$th path estimated by the BS in this case is the mirror position of the UE concerning the reflection plane, denoted as Virtual UE (vUE). We determine the real position of the scatterer by the geometrical relationship between the vUE, UE, and BS.

We denote the positions of the real scatterer, vUE, UE, and BS as ${{\textbf{x}}_{l}}=\left[ {{x}_{l}},{{y}_{l}},{{z}_{l}} \right]$, ${{\textbf{x}}_{vl}}=\left[ {{x}_{vl}},{{y}_{vl}},{{z}_{vl}} \right]$, ${{\textbf{x}}_{1}}=\left[ {{x}_{1}},{{y}_{1}},{{z}_{1}} \right]$, and ${{\textbf{x}}_{0}}=\left[ {{x}_{0}},{{y}_{0}},{{z}_{0}} \right]$, respectively. Then the midpoint of UE and vUE is ${{\text{M}}_{1,vl}}=\left[ \frac{{{x}_{1}}+{{x}_{vl}}}{2},\frac{{{y}_{1}}+{{y}_{vl}}}{2},\frac{{{z}_{1}}+{{z}_{vl}}}{2} \right]$. The normal vector ${{\vec{u}}_{1,vl}}$ of a line segment in space consisting of UE and vUE is defined as
\begin{equation}
    {{\vec{u}}_{1,vl}}=\frac{\left[ {{x}_{vl}}-{{x}_{1}},{{y}_{vl}}-{{y}_{1}},{{z}_{vl}}-{{z}_{1}} \right]}{\left| \left[ {{x}_{vl}}-{{x}_{1}},{{y}_{vl}}-{{y}_{1}},{{z}_{vl}}-{{z}_{1}} \right] \right|}
    \label{eq:u_1_vl}
\end{equation}
Substituting ${{\text{M}}_{1,vl}}$ into (\ref{eq:u_1_vl}) gives the equation of the perpendicular plane of the above line segment
\begin{equation}
    {{\vec{u}}_{1,vl}}\cdot \left( \vec{r}-{{\text{M}}_{1,vl}} \right)=0
    \label{eq:perpendicular}
\end{equation}
The coordinate of the intersection of the line where BS and vUE are located and the above plane is the true position of the scatterer. The normal vector of the line between BS and vUE is given by
\begin{equation}
    {{\vec{u}}_{0,vl}}=\frac{\left[ {{x}_{vl}}-{{x}_{0}},{{y}_{vl}}-{{y}_{0}},{{z}_{vl}}-{{z}_{0}} \right]}{\left| \left[ {{x}_{vl}}-{{x}_{0}},{{y}_{vl}}-{{y}_{0}},{{z}_{vl}}-{{z}_{0}} \right] \right|}
    \label{eq:u_0_vl}
\end{equation}
The parametric equation of the above line with unknown parameter $t$ after substituting ${{\textbf{x}}_{0}}$ is 
defined as
\begin{equation}
    {{\vec{Q}}_{0,l}}={{\textbf{x}}_{0}}+t{{\vec{u}}_{0,vl}}
    \label{eq:parametric_equation}
\end{equation}
Joining (\ref{eq:u_1_vl}) and (\ref{eq:parametric_equation}) gives
\begin{equation}
    {{\vec{u}}_{0,vl}}\cdot \left( \left( {{\text{x}}_{0}}+t{{{\vec{u}}}_{0,vl}} \right)-{{\text{M}}_{1,vl}} \right)=0
    \label{eq:lasteq}
\end{equation}
We can calculate the unknown parameter $t$ by substituting the relevant data into (\ref{eq:lasteq}). And the position information of the real scatterer can be solved by applying $t$ into (\ref{eq:parametric_equation}).

\section{Methodology of Sensing Result Fusion and Enhancement} \label{section4}
In this section, we demonstrate the fusion method of DL-UL ISAC 4D environmental reconstruction results. Firstly, we show the data-level fusion strategy based on the simple solution of the theory of Sphere Packings; then we show the AGGRNN network model based on deep learning for improving the point cloud density of the environmental reconstruction results after data-level fusion; and finally, we present the feature-level fusion reconstruction network MVSFNet based on the multi-view feature-level fusion strategy. 


\subsection{Data Level Fusion Strategy} \label{datafusion}
\begin{figure}[!htb]
	\centering  
	\subfigure[Original reconstruction result.]{
		\includegraphics[width=0.45\linewidth]{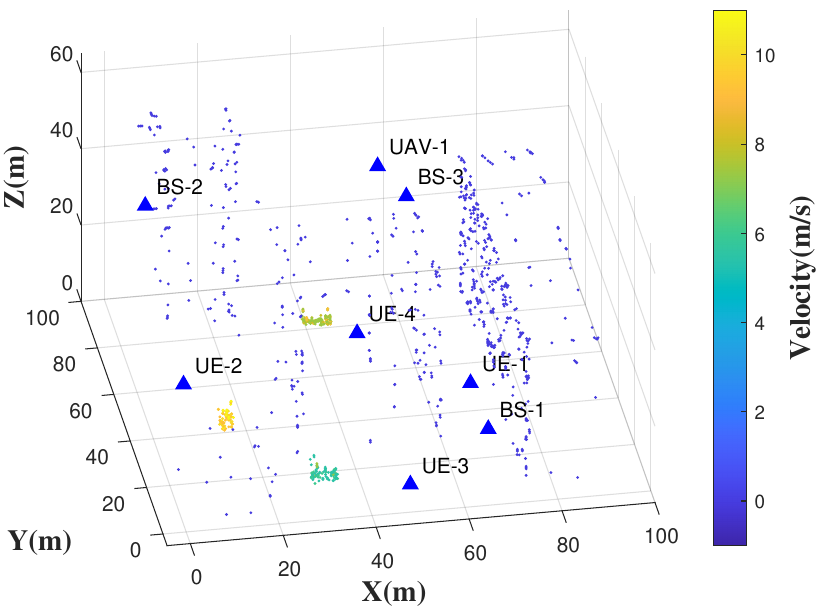}}
 	\subfigure[Data-level fusion result.]{
		\includegraphics[width=0.45\linewidth]{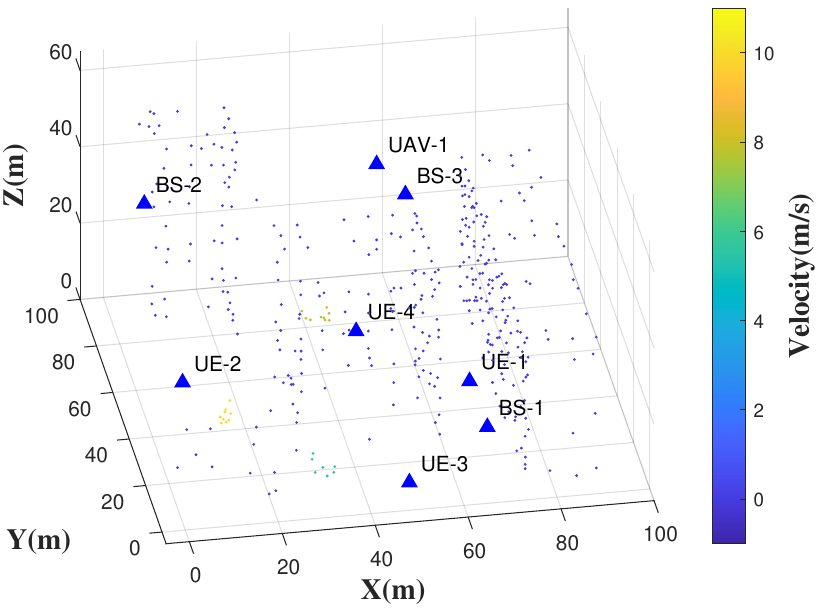}}
	\caption{(a) and (b) are the result of the original ISAC 4D environmental reconstruction and the data-level fusion result, respectively.}
	\label{fig:datalevel_fusion}
\end{figure}

\subsubsection{Cartesian Coordinate System Rotation and Unification}
In order to realize the environmental reconstruction, it is necessary to restore their coordinates to the world coordinate system from  the BS coordinate system by using the Eulerian rotation operators ${{\textbf{R}}_{ZYX}}$.



\begin{equation}
    \footnotesize
    \begin{aligned}
      & {{\textbf{R}}_{ZYX}}={{\textbf{R}}_{Z}}(\gamma ){{\textbf{R}}_{Y}}(\beta ){{\textbf{R}}_{X}}(\alpha ) \\ 
     & =\left[ \begin{matrix}
       \cos \gamma  & -\sin \gamma  & 0  \\
       \sin \gamma  & \cos \gamma  & 0  \\
       0 & 0 & 1  \\
    \end{matrix} \right]\left[ \begin{matrix}
       \cos \beta  & 0 & \sin \beta   \\
       0 & 1 & 0  \\
       -\sin \beta  & 0 & \cos \beta   \\
    \end{matrix} \right]\left[ \begin{matrix}
       1 & 0 & 0  \\
       0 & \cos \alpha  & -\sin \alpha   \\
       0 & \sin \alpha  & \cos \alpha   \\
    \end{matrix} \right] \\ 
    \end{aligned}
    \label{eq:rz}
\end{equation}

The coordinates of a certain scatterer in the world coordinate system is given by
\begin{equation}
    {{\textbf{P}}_{\text{World}}}={{\textbf{P}}_{\text{BS}}}\times {{\textbf{R}}_{ZYX}}+{{\textbf{P}}_{\text{Offset}}}
    \label{eq:rz}
\end{equation}
where ${{\text{P}}_{\text{Offset}}}=\left[ {{x}_{o}},{{y}_{o}},{{z}_{o}} \right]$ is the positional offset of the BS with respect to the reference point of the world coordinate system, ${{\text{P}}_{\text{BS}}}=\left[ x,y,z \right]$ is the scatterer coordinates of the BS coordinate system, and ${{\text{P}}_{\text{World}}}=\left[ {x}',{y}',{z}' \right]$ is the scatterer coordinates of the transformed world coordinate system; and $\alpha$, $\beta$, and $\gamma$ are the angles of rotation of the BS coordinate system around the $x$, $y$, and $z$ axes, respectively.

\subsubsection{Offset Scatterers Fusion}
The ISAC 4D environmental reconstruction results after directly stitching the sensing results from multiple nodes are shown in Fig.~\ref{fig:datalevel_fusion}(a). It can be seen that the coordinates of the same scatterer are slightly shifted. The key to data-level fusion is to select the appropriate fusion radius $R_c$.
\begin{figure}[!htb]
	\centering
	\includegraphics[width=0.55\linewidth]{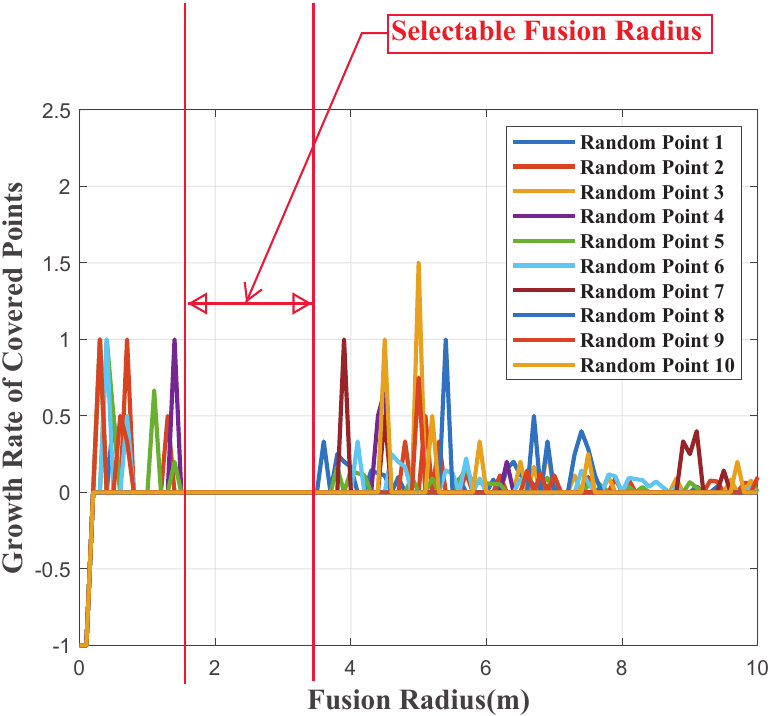}
	\caption{Schematic of the relationship between the fusion radius of the covered sphere and the growth rate of the number of points in the sphere.}
	\label{fig:covered_sphere}
\end{figure}

We randomly select 10 scatterers and create expanding covering spheres with radius $R_c$, and we define the growth rate ${h}_{i}$ of the number of scatterers in the covering sphere as a monitoring metric as
\begin{equation}
    {{h}_{i}}=\frac{{{C}_{n}}-{{C}_{b}}}{{{C}_{b}}}
    \label{eq:rate_increase}
\end{equation}
where ${C}_{n}$ is the number of scatterers in the current coverage sphere and ${C}_{n}$ is the number of scatterers in the coverage sphere at the previous differential radius. We calculate the ${h}_{i}$ variation curve when ${{R}_{c}}\in \left[ 0,10 \right]$ as shown in Fig.~\ref{fig:covered_sphere}.

Fig.~\ref{fig:covered_sphere} shows that this interval is an selectable interval for the fusion radius $R_c$ when the radius of all covered spheres increases but the growth rate of the number of scatterers in the sphere satisfied ${{h}_{i}}\to 0$.

\subsection{AGGRNN: Low-density Reconstruction Point Cloud Enhancing Method}
\begin{figure*}[!htb]
	\centering
	\includegraphics[width=0.75\linewidth]{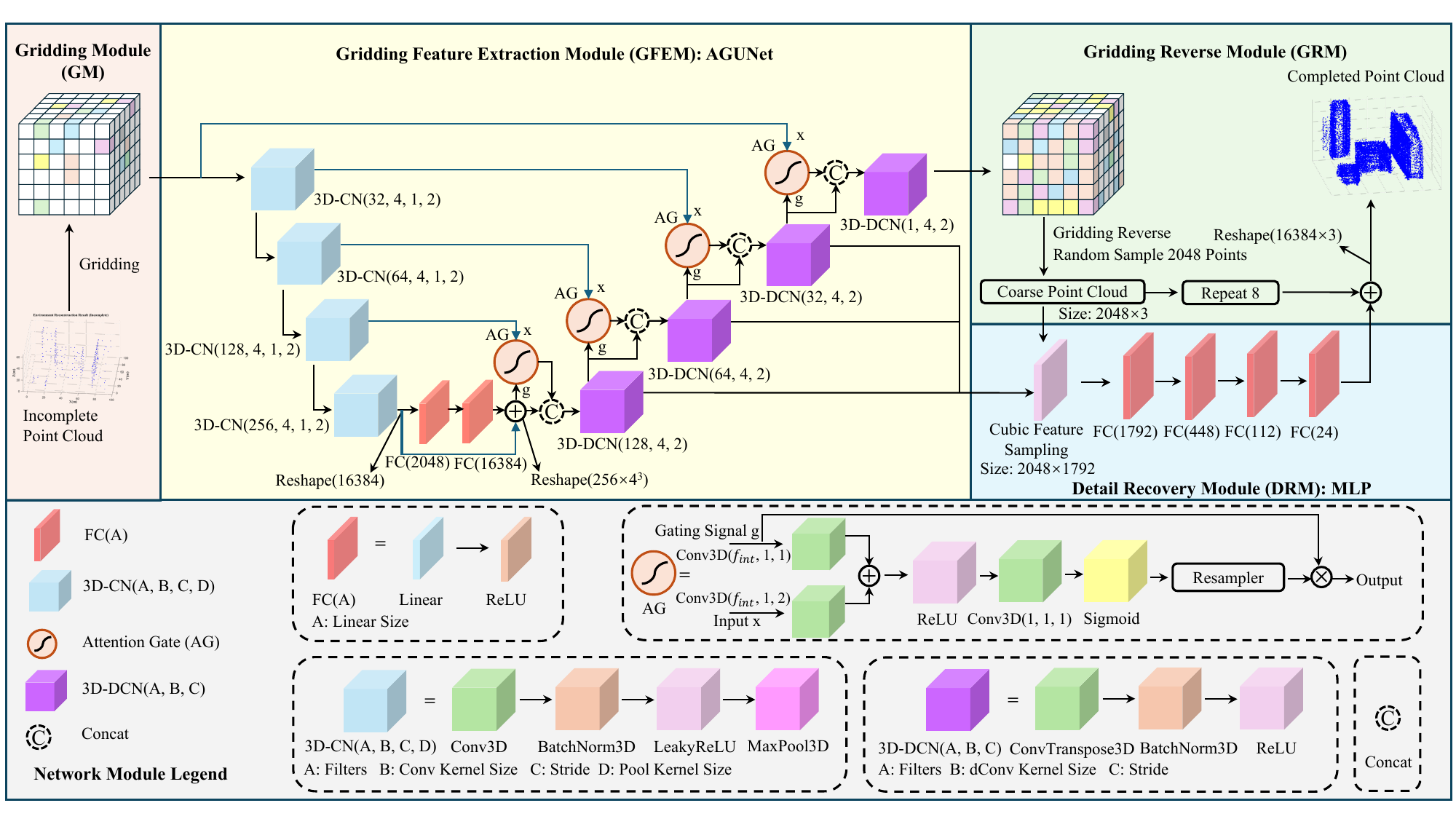}
	\caption{Overall framework of AGGRNN.}
	\label{fig:AGGRNN}
\end{figure*}
As shown in Fig.~\ref{fig:AGGRNN}, AGGRNN contains four components are Gridding module (GM), Gridding Feature Extraction Module (GFEM), Detail Recovery Module (DRM) and Gridding Reverse Module (GRM). GM first transforms the input low-density point cloud into a 3D grid $\xi =\left\langle V,K \right\rangle $, where $V$ and $K$ are the vertex set and value set of $\xi$, respectively. Then GFEM extracts $\xi$'s high dimensional spatial features output as ${K}'$ through AGUNet. The DRM then works with the GRM to generate the spatial features $F^{c}$ of the coarse point cloud $P^{c}$ and uses the Multi-Layer Perceptron (MLP) to generate the adjustment vector $W^{c}$ of $P^{c}$. Finally, GRM superimposes the adjustment vector $W^{c}$ on the generated coarse point cloud $P^{c}$ to output the high-density point cloud $P^{f}$\cite{xie2020grnet}. The Loss function for our supervised model training is $Gridding Loss$.
\begin{figure}[!htb]
	\centering  
	\subfigbottomskip=2pt 
	\subfigcapskip=-5pt 
	\subfigure[]{
		\includegraphics[width=0.2\linewidth]{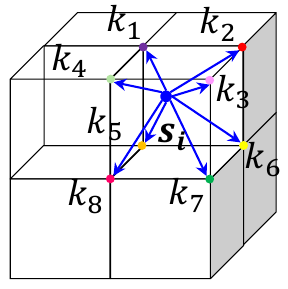}}
 	\subfigure[]{
		\includegraphics[width=0.2\linewidth]{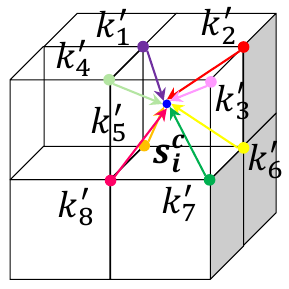}}
	\caption{(a) and (b) are Gridding and Gridding Reverse, respectively.}
	\label{fig:G&GR}
\end{figure}
\subsubsection{Gridding Module (GM)}
In GM we introduce 3D grids as intermediate representations of regularized point clouds, and further introduce a differentiable Gridding layer that converts a disordered, irregular point cloud $P=\left\{ {{p}_{i}} \right\}_{i=1}^{n}$ into an ordered, regular point cloud $\xi =\left\langle V,K \right\rangle $ while preserving the spatial layout of the point cloud, where ${{p}_{i}}\in {{\mathbb{R}}^{3}}$, $V=\left\{ {{v}_{i}} \right\}_{i=1}^{{{N}^{3}}}$, $K=\left\{ {{k}_{i}} \right\}_{i=1}^{{{N}^{3}}}$, ${{v}_{i}}\in \left\{ \left( -\frac{N}{2},-\frac{N}{2},-\frac{N}{2} \right),...,\left( -\frac{N}{2}-1,-\frac{N}{2}-1,-\frac{N}{2}-1 \right) \right\}$, ${{k}_{i}}\in \mathbb{R}$, $n$ is the number of points in $P$, and $N$ is the resolution of the 3D grid $\xi$. As shown in Fig.~\ref{fig:G&GR}(a), the 3D Grid contains a number of Cells, and each Cell contains 8 vertices, for each vertex ${{v}_{i}}=\left( x_{i}^{v},y_{i}^{v},z_{i}^{v} \right)$ in the Cell, if the point $p=\left( x,y,z \right)$ in the low-density point cloud satisfies \ref{eq:p_requirement},then $p$ falls in the neighborhood $\mathcal{N}$ of vertex $v_i$, denoted as $p\in \mathcal{N}\left( {{v}_{i}} \right)$.
\begin{equation}
\footnotesize
    \left\{ \begin{aligned}
      & x_{i}^{v}-1<x<x_{i}^{v}+1 \\ 
     & y_{i}^{v}-1<y<y_{i}^{v}+1 \\ 
     & z_{i}^{v}-1<z<z_{i}^{v}+1 \\ 
    \end{aligned} \right.
    \label{eq:p_requirement}
\end{equation}
As shown in Fig.~\ref{fig:G&GR}(a), for a given vertex $v_i$ with $\mathcal{N}$, the corresponding weight $k_i$ of the vertex is defined as
\begin{equation}
    {{k}_{i}}=\sum\limits_{p\in \mathcal{N}\left( {{v}_{i}} \right)}{\frac{k\left( {{v}_{i}},p \right)}{\left| \mathcal{N}\left( {{v}_{i}} \right) \right|}}
    \label{eq:ki}
\end{equation}
where ${{k}_{i}}\left( {{v}_{i}},p \right)=\left( 1-\left| x_{i}^{v}-x \right| \right)\left( 1-\left| y_{i}^{v}-y \right| \right)\left( 1-\left| z_{i}^{v}-z \right| \right)$, in particular, for$\left| \mathcal{N}\left( {{v}_{i}} \right) \right|=0,{{k}_{i}}=0$.

\subsubsection{Gridding Feature Extraction Module (GFEM)}
GFEM aims to extract the depth features of low-density point clouds in terms of spatial arrangement and spatial distribution. Its basic structure is AGUNet, which is a 3D encoder-decoder structure introducing AG modules with U-net connections\cite{cciccek20163d}. The details of the network parameters and network structure are shown in Fig.~\ref{fig:AGGRNN}. For a given set $K$ of values of a 3D Grid, the GFEM process can be given formally as ${K}'=\text{GFEM}\left( K \right)$, 
where ${K}'=\left\{ {{{{k}'}}_{i}} \right\}_{i=1}^{{{N}^{3}}},{k}'\in \mathbb{R}$.

\subsubsection{Detail Recovery Module(DRM)}\label{drm}
DRM is composed of Cubic Feature Sampling (CFS) layer and MLP. We introduce a CFS layer\cite{xie2020grnet} to extract features at different depths of the point cloud from the up-sampled 3D feature maps and preserve the contextual information. 

\begin{figure}[!htb]
	\centering
	\includegraphics[width=0.65\linewidth]{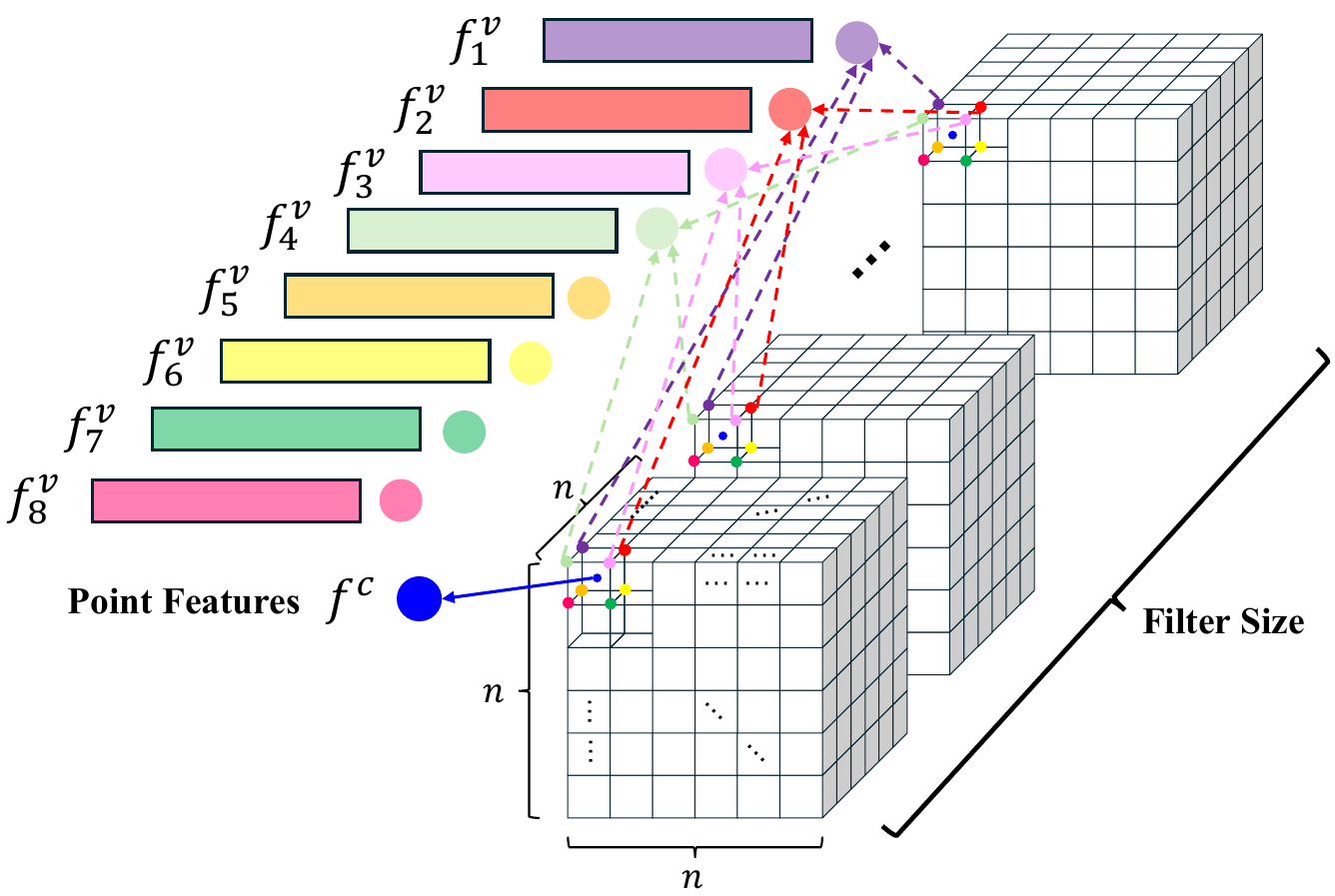}
	\caption{Overall framework of CFS layer.}
	\label{fig:CFS}
\end{figure}

As shown in Fig.~\ref{fig:CFS}, the feature set $\mathcal{F}=\left\{ \left. f_{1}^{v},f_{2}^{v},...,f_{{{t}^{3}}}^{v} \right|f_{i}^{v}\in {{\mathbb{R}}^{c}} \right\}$ of a certain feature map of AGUNet and a certain point $p_{i}^{c}$ in the coarse point cloud $P^c$ are given, and the feature $f_{i}^{c}$ corresponding to $p_{i}^{c}$ can be denoted as $f_{i}^{c}=\left[ f_{\theta _{1}^{i}}^{v},f_{\theta _{2}^{i}}^{v},...,f_{\theta _{8}^{i}}^{v} \right]$, 
where $\left[ \cdot  \right]$ is the concatenation operation, $\left\{ f_{\theta _{j}^{i}}^{v} \right\}_{j=1}^{8}$denotes the set of features of the eight vertices of the Cell where $p_{i}^{c}$ is located, the coarse point cloud $P^c$ is obtained from the GRM in Section \ref{GRM}. We randomly sample 2048 points from the coarse point cloud $P^c$, and the CFS layer extracts features from the three up-sampled feature maps of AGUNet and generates the feature $F^c$ of the above 2048 points, as shown in Fig.~\ref{fig:CFS}.

The MLP estimates $r$ offsets for each point $p_{i}^{c}$ to generate the final point cloud. Specifically, given a coarse point cloud $P^c$ with its corresponding feature $F^c$, the MLP and GM eventually collaborate to generate a high-density point cloud ${{P}^{f}}=\left\{ p_{i}^{f} \right\}_{i=1}^{n}$.
\begin{equation}
    {{P}^{f}}=\text{MLP}\left( {{F}^{c}} \right)+\text{GM}\left( {{P}^{c}},r \right)
    \label{eq:pf}
\end{equation}
where $p_{i}^{f}\in {{\mathbb{R}}^{3}}$, $n$ is the number of points in the final high-density point cloud ${P}^{f}$. GM creates a new tensor by replicating $P^c$ $r$ times. As shown in Fig.~\ref{fig:AGGRNN}, we take $r=8$ and the output of MLP is reshaped to $r\times m \times 3=16384 \times 3$, which corresponds to the offsets of the coordinates of $16384$ points.
\subsubsection{Gridding Reverse Module (GRM)} \label{GRM}
GM consists of $Gridding Reverse$ with tail processing of AGGRNN. $Gridding Reverse$ is the inverse of Gridding which converts the 3D Grid back to a point cloud. $Gridding Reverse$ takes the 3D Grid ${\xi }'=\left\langle V,{K}' \right\rangle $ generated by AGUNet as input to generate a coarse point cloud ${{P}^{c}}=\left\{ p_{i}^{c} \right\}_{i=1}^{m}$, where $p_{i}^{c}\in {{\mathbb{R}}^{3}}$ and $m$ is the number of points in the coarse point cloud ${{P}^{c}}$. As shown in Fig.~\ref{fig:G&GR}(b), $Gridding Reverse$ generates a point $p_{i}^{c}$ for each Cell in ${\xi }'$, whose coordinates can be obtained by weighting the coordinates of the 8 vertices of the current Cell with the weights.
\begin{equation}
    p_{i}^{c}=\frac{\sum\nolimits_{\theta \in {{\Theta }^{i}}}{{{{{w}'}}_{\theta }}{{v}_{\theta }}}}{\sum\nolimits_{\theta \in {{\Theta }^{i}}}{{{{{w}'}}_{\theta }}}}
    \label{eq:GR}
\end{equation}
where ${{\Theta }^{i}}=\left\{ \theta _{j}^{i} \right\}_{j=1}^{8}$ represents the set of subscript indexes of the vertices of this Cell, $\left\{ \left. {{v}_{\theta }} \right|\theta \in {{\Theta }^{i}} \right\}$ and $\left\{ \left. {{{{w}'}}_{\theta }} \right|\theta \in {{\Theta }^{i}} \right\}$ represent the set of vertices and values of the Cell, respectively. In particular, if $\sum\nolimits_{\theta \in {{\Theta }^{i}}}{{{{{w}'}}_{\theta }}}=0$, then $Gridding Reverse$ will not generate a point $p_{i}^{c}$ for the Cell. The tail processing section is described in Section \ref{drm} and will not be repeated here.

\subsubsection{Gridding Loss}
\begin{figure}[!htb]
	\centering
	\includegraphics[width=0.7\linewidth]{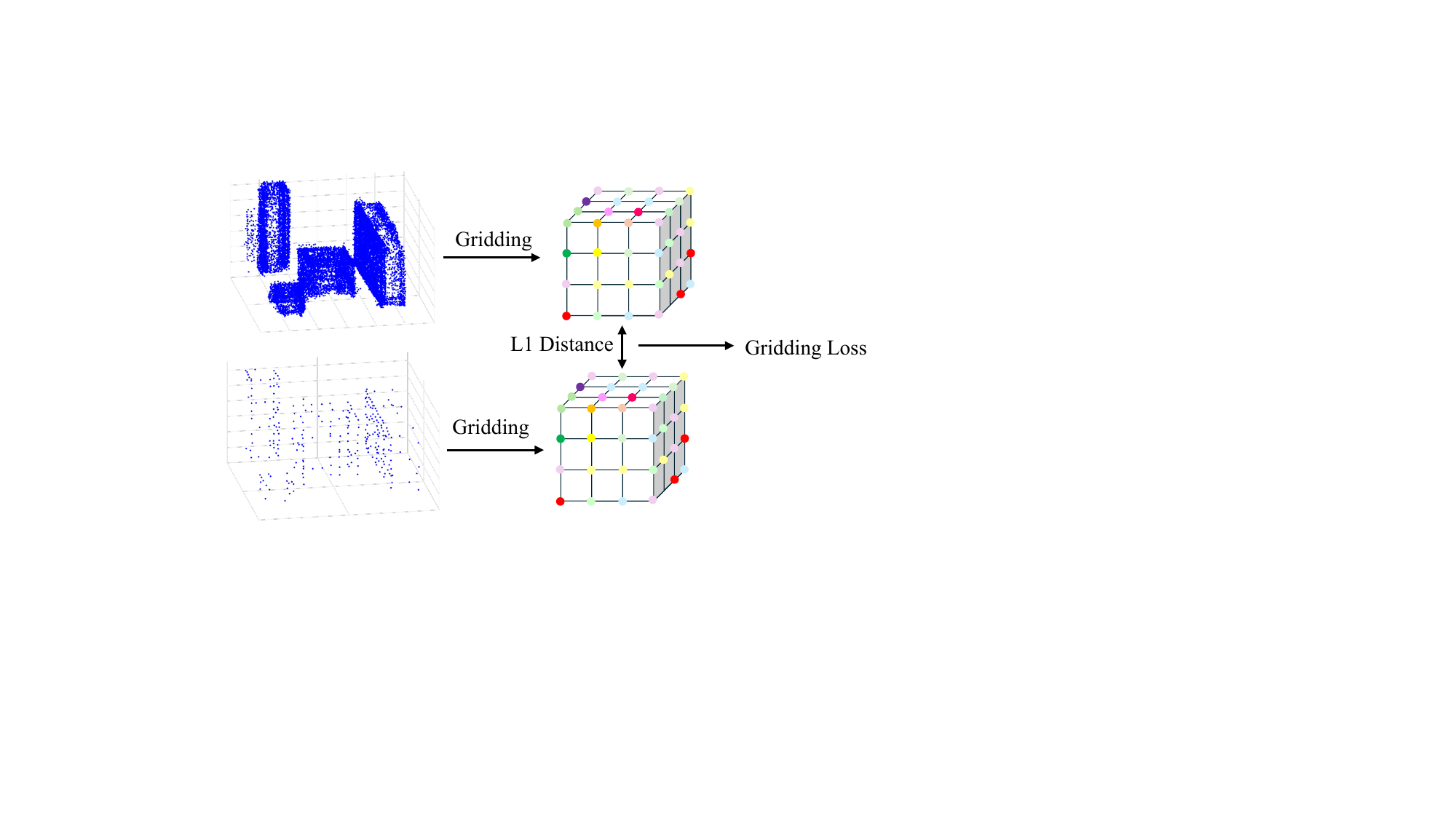}
	\caption{Gridding Loss.}
	\label{fig:Gloss}
\end{figure}
We introduce a new loss function $Gridding Loss$ based on GM, defined as the L1 distance between two 3D Grids value sets ${{\xi }_{Pred}}=\left\langle {{V}^{Pred}},{{K}^{Pred}} \right\rangle$ and ${{\xi }_{GT}}=\left\langle {{V}^{GT}},{{K}^{GT}} \right\rangle$, respectively\cite{xie2020grnet}.
\begin{equation}
    {{\mathcal{L}}_{Gridding}}\left( {{K}^{Pred}},{{K}^{GT}} \right)=\frac{1}{N_{G}^{3}}\sum{\left\| {{K}^{Pred}}-{{K}^{GT}} \right\|}
    \label{eq:Gloss}
\end{equation}
Where ${{K}^{Pred}}\in {{\mathbb{R}}^{N_{G}^{3}}},{{K}^{GT}}\in {{\mathbb{R}}^{N_{G}^{3}}}$, $N_G$ is the resolution of the two 3D Grids.

\subsection{MVSFNet: Feature Level Fusion Strategy}
\begin{figure*}[!htb]
	\centering
	\includegraphics[width=0.75\linewidth]{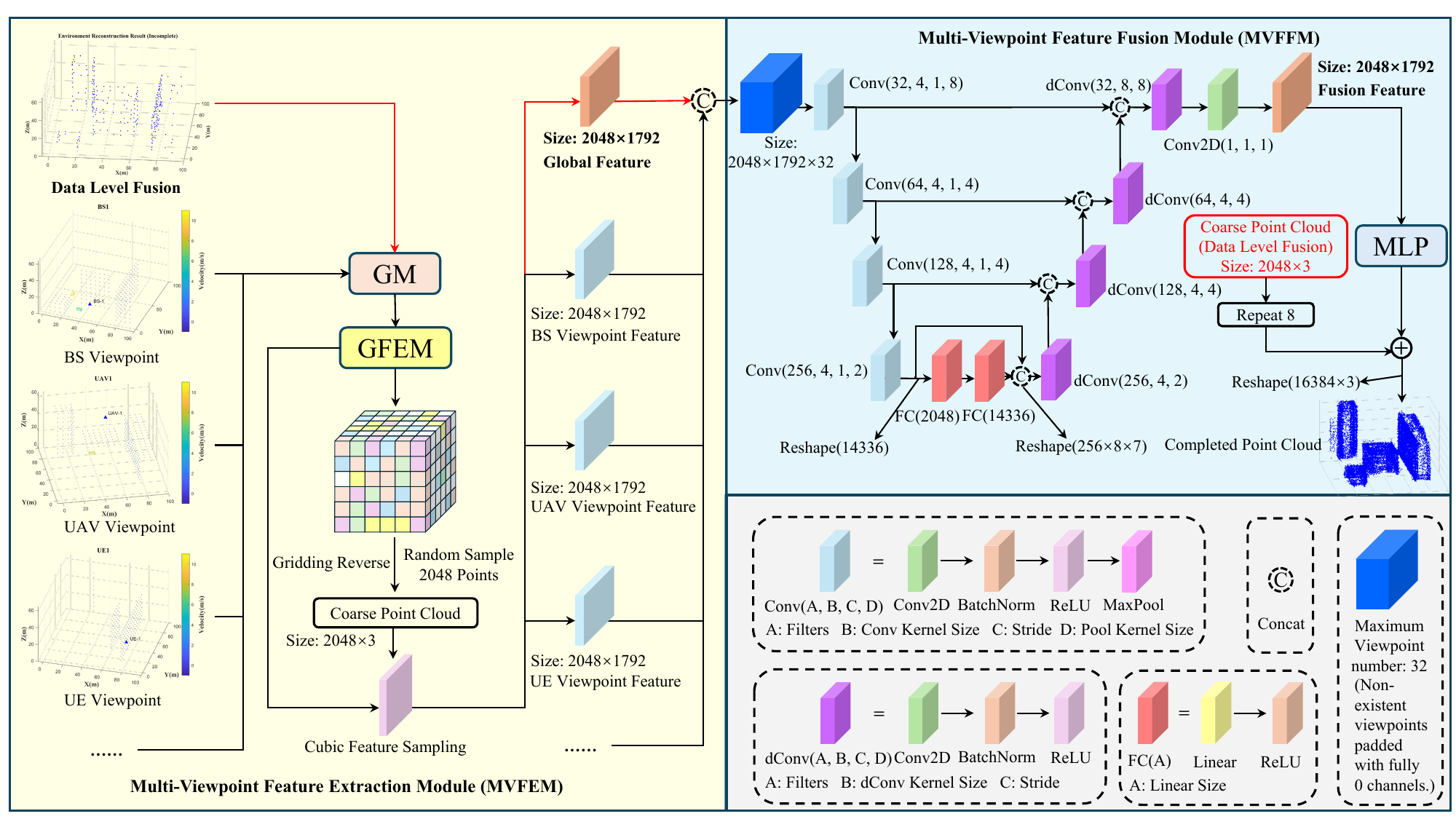}
	\caption{Overall framework of MVSFNet.}
	\label{fig:MVSFNet}
\end{figure*}
As shown in Fig.~\ref{fig:MVSFNet}, MVSFNet consists of two parts, Multi-Viewpoint Feature Extraction Module (MVFEM) and Multi-Viewpoint Feature Fusion Module (MVFFM). 

\subsubsection{Multi-Viewpoint Feature Extraction Module (MVFEM)} \label{MVFEM}
The coarse point cloud set $\mathcal{P}=\left\{ P_{g}^{c},P_{1}^{c},...,P_{31}^{c} \right\}$ of reconstruction results from different viewpoints can be obtained after GM, GFEM, and $Gridding Reverse$, where $P_{g}^{c}$ is the coarse point cloud generated from the global low-density point cloud after data-level fusion, $P_{\eta }^{c}\left( \eta \in \left[ 1,31 \right] \right)$ is the coarse point cloud generated from a multi-view low-density point cloud. $F_{full}^{c}=\left[ F_{g}^{c},F_{1}^{c},...,F_{31}^{c} \right]$, 
where $\left[ \cdot  \right]$ is the concatenation operation, $F_{g}^{c}$ corresponds to the feature map of global low-density point cloud $P_{g}^{c}$ after data-level fusion, $F_{\eta }^{c}\left( \eta \in \left[ 1,31 \right] \right)$ corresponds to the feature map of  multi-view low-density point cloud. 

We specify the maximum number of channels of $F_{full}^{c}$ to be 32, so the maximum number of viewpoints supported by MVSFNet is 31 after removing the stationary-occupied global feature map $F_{g}^{c}$. If the number of viewpoints is less than 31, the corresponding $F_{\eta }^{c}\left( \eta \in \left[ 1,31 \right] \right)$ adopts an all-zero Tensor. 

\subsubsection{Multi-Viewpoint Feature Fusion Module (MVFFM)}
 We use a two-dimensional encoder-decoder structure for fusing multi-view feature maps in MVFEM, replacing the stationary feature map weights by trainable network parameters. The specific network parameters and network structure of MVFFM are shown in Fig.~\ref{fig:MVSFNet}. It is worth noting that the point cloud that is complemented in the tail processing is still the coarse point cloud $P_{g}^{c}$ from the global low-density point cloud.

\section{Simulation Results and Analysis} \label{sec5}
This section first introduces the evaluation metrics of MNDUC ISAC 4D environmental reconstruction method; then introduces the methods of generating the training and evaluation data; then introduces the parameter settings of the simulation; and finally, the qualitative and quantitative evaluations of the MNDUC ISAC 4D environmental reconstruction method proposed in this research are conducted.
\subsection{Dataset Generation}
\begin{figure}[!htb]
	\centering  
	\subfigure[Simulation scenario modeling.]{
		\includegraphics[width=0.4\linewidth]{figure/abstractenvironment3D.pdf}}
  	\subfigure[Sampling of scatterers.]{
		\includegraphics[width=0.4\linewidth]{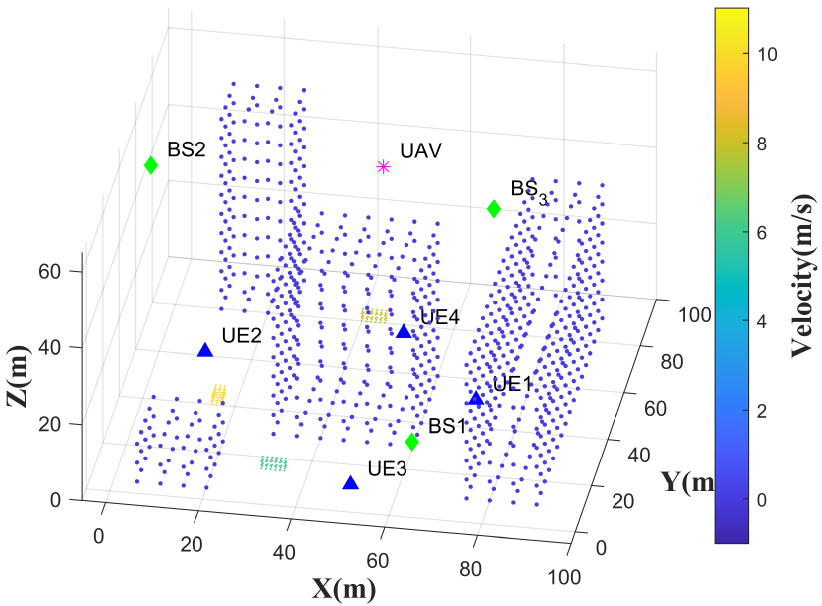}}
  \\
	\subfigure[MNDUC sensing.]{
		\includegraphics[width=0.4\linewidth]{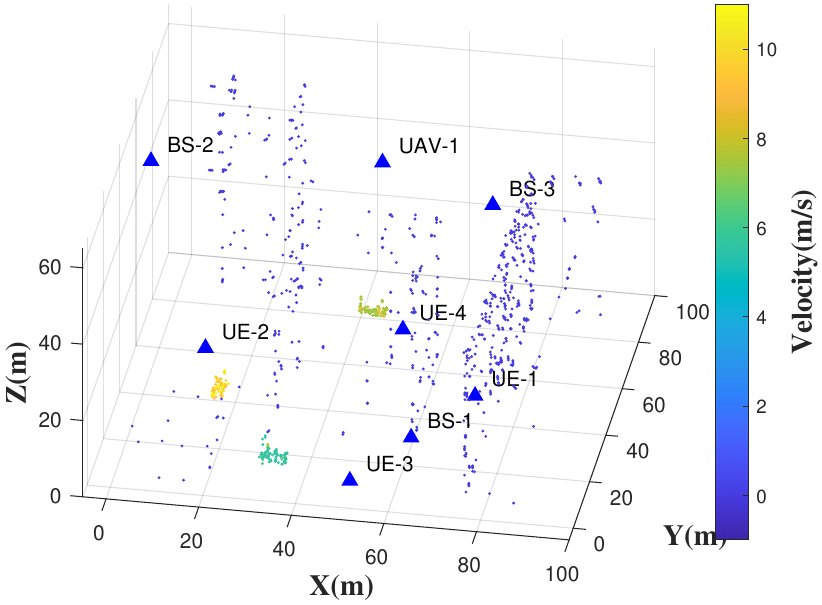}}
  	\subfigure[Data-level fusion.]{
		\includegraphics[width=0.4\linewidth]{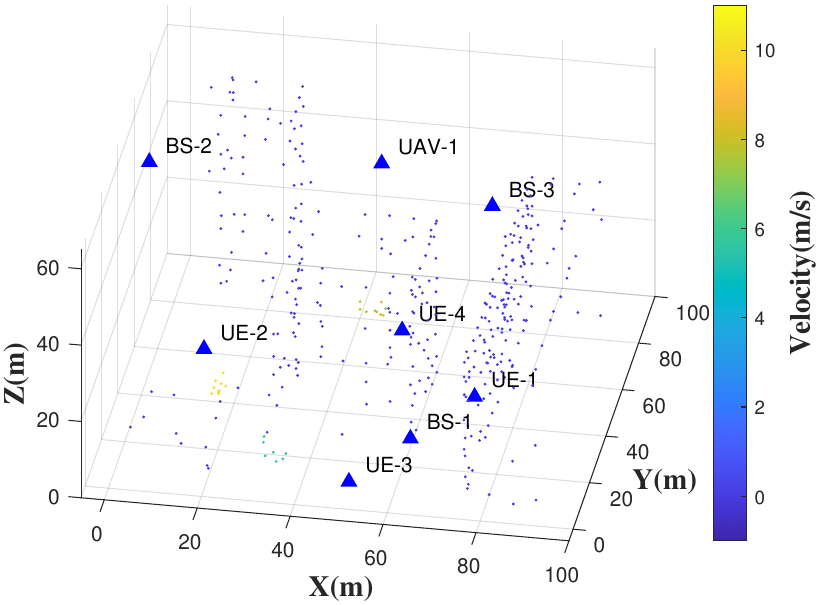}}
	\caption{(a)-(d) show the generation process of our dataset.}
	\label{fig:datasetflow}
 \end{figure}
 As shown in Fig.~\ref{fig:datasetflow}(a), we first randomly generate stationary buildings and moving vehicles of different sizes but not overlapping each other in a fixed-size 3D space, and then randomly place sensing nodes such as BSs, UEs, UAVs, and so on, in the areas that are not covered by the buildings and vehicles. As shown in Fig.~\ref{fig:datasetflow}(b), we construct a set of scatterers in the scenario space by extracting scattering points on the surfaces of buildings and vehicles. As shown in Fig.~\ref{fig:multi-view}(a)(c)(e), we calculate the scatterers locate on the LoS paths and single reflection NLoS paths in DL and UL period of different sensing nodes by Raytracing and occlusion detection algorithms. The environmental reconstruction results of the corresponding selected sensing nodes are shown in Fig.~\ref{fig:multi-view}(b)(d)(f), and their direct fusion results are shown in Fig.~\ref{fig:datasetflow}(c). As shown in Fig.~\ref{fig:datasetflow}(d), we apply the data-level fusion strategy in Section \ref{datafusion} to process the direct fusion results to obtain a low-density reconstructed point cloud as the training data for AGGRNN. For the Multi-Modal Model MVSFNet, one of its inputs is the low-density point cloud training data of the AGGRNN described above, and the other input is the feature maps extracted from the multi-view sensing result, and the extraction method is described in Section \ref{MVFEM}.

 The specifications of this dataset are 10,000 data for the training set, 100 data for the validation set, and 100 data for the test set. And it is labeled in ShapeNet format.

\subsection{Simulation Parameters and Environment Setting}

\begin{table}[!htb]
    \centering
    \caption{Simulation Parameters of System for DL and UL Periods.}
    \label{table:para}
    \setlength{\tabcolsep}{0.7mm} 
    \begin{tabular}{@{}c|cc@{}}
    \toprule
    \textbf{Parameter names}                                                                  & \textbf{DL}                                                                             & \textbf{UL}                  \\ \midrule
    MIMO size                                                                                 & \begin{tabular}[c]{@{}c@{}}$8\times 8$ Tx, $2\times 2$ Rx\\ $16\times 16$ vRx\end{tabular} & $1\times 1$ Tx, $8\times 8$ Rx \\
    Carrier frequency                                                                         & 70 GHz                                                                                   & 70 GHz                        \\
    Bandwidth                                                                                 & 491.52 MHz                                                                               & 209.76 MHz                    \\
    Subcarrier spacing                                                                        & 240 kHz                                                                                  & 240 kHz                       \\
    Subcarrier count                                                                          & 2048                                                                                    & 1024                         \\
    OFDM symbol count                                                                         & 224                                                                                     & 224                          \\
    Slot count                                                                                & 16                                                                                      & 16                           \\
    Spatially smoothed subarray size                                                          & 8                                                                                       & 4                            \\ \bottomrule
    \end{tabular}
\end{table}
The signal and array parameters in the DL period and the UL period are set as shown in Table~\ref{table:para}. 

The experiments are all implemented on an $\textrm{Intel}^@$ $\textrm{Core}^{TM}$ i9-13900KF CPU (4.3GHz, 64GB RAM) and an NVIDIA GeForce RTX 4090 GPU (CUDA version 11.6) with Python 3.8.18 (PyTorch 2.1.2) in 64 Bit Ubuntu 22.04.1 Long Term Support operating system. We select the Adaptive Moment Estimation (Adam) optimizer with ${\beta}_1=0.9, {\beta}_2=0.99$ for finding the optimal parameters, the initial learning rate is 0.00001 and the $Batchsize$ is set to 32.

\subsection{Evaluation metrics}
In our experiments, we use both Chamfer
Distance (CD) and F-Score as quantitative evaluation metrics.
\subsubsection{Chamfer Distance}
The distance between $\mathcal{T}$ and $\mathcal{R}$ are defined as (\ref{eq:CD})\cite{tchapmi2019topnet,fan2017point}
\begin{equation}
    {{\mathcal{D}}_{\text{CD}}}=\frac{1}{\left| \mathcal{T} \right|}\sum\limits_{t\in \mathcal{T}}{\underset{r\in \mathcal{R}}{\mathop{\min }}\,\left\| t-r \right\|_{2}^{2}}+\frac{1}{\left| \mathcal{R} \right|}\sum\limits_{r\in \mathcal{R}}{\underset{t\in \mathcal{T}}{\mathop{\min }}\,\left\| t-r \right\|_{2}^{2}}
    \label{eq:CD}
\end{equation}
where $\mathcal{T}=\left\{ \left( {{x}_{i}},{{y}_{i}},{{z}_{i}} \right) \right\}_{i=1}^{{{n}_{\mathcal{T}}}}$ is the Ground Truth(GT) and $\mathcal{R}=\left\{ \left( {{x}_{i}},{{y}_{i}},{{z}_{i}} \right) \right\}_{i=1}^{{{n}_{\mathcal{R}}}}$ is the reconstructed point set being evaluated, ${{{n}_{\mathcal{T}}}}$ and ${{{n}_{\mathcal{R}}}}$ are the numbers
of points in $\mathcal{T}$ and $\mathcal{R}$, respectively.
\subsubsection{F-Score}
We introduce F-Score as an extended metric to evaluate the performance of low-density point cloud enhancement networks, which can be defined as\cite{xie2020grnet}
\begin{equation}
    \text{F-Score}\left( d \right)=\frac{2P\left( d \right)R\left( d \right)}{P\left( d \right)+R\left( d \right)}
    \label{eq:Fscore}
\end{equation}
where $P\left( d \right)=\frac{1}{{{n}_{\mathcal{R}}}}\sum\limits_{r\in \mathcal{R}}{\left[ \underset{t\in \mathcal{T}}{\mathop{\min }}\,\left\| t-r \right\|<d \right]}$ and $R\left( d \right)=\frac{1}{{{n}_{\mathcal{T}}}}\sum\limits_{t\in \mathcal{T}}{\left[ \underset{r\in \mathcal{R}}{\mathop{\min }}\,\left\| t-r \right\|<d \right]}$ are the precision and recall for a distance threshold $d$, respectively.

\subsection{Qualitative Analysis of Simulation Results}
\subsubsection{Comparison of Single Node and Multi-Node Cooperative ISAC 4D Environmental Reconstruction Results}
\begin{figure}[!htb]
	\centering  
	\subfigure[DL, Original, BS1.]{
		\includegraphics[width=0.4\linewidth]{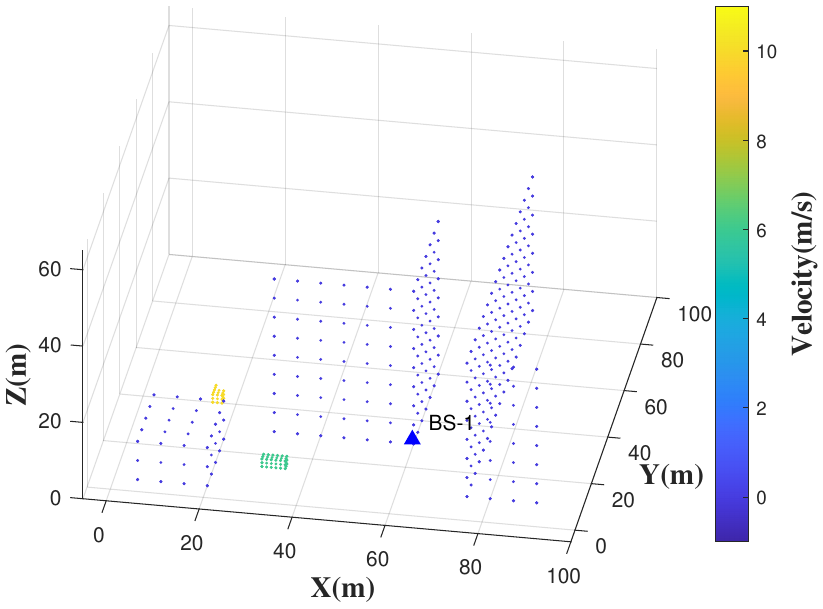}}
	\subfigure[DL, Rresult, BS1.]{
		\includegraphics[width=0.4\linewidth]{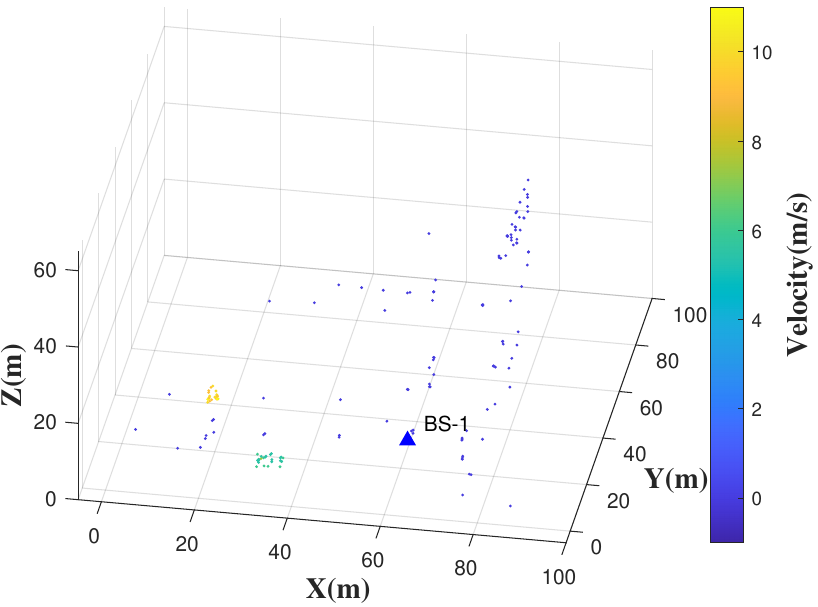}}
  \\
  	\subfigure[UL, Original, UE1 to BS2.]{
		\includegraphics[width=0.4\linewidth]{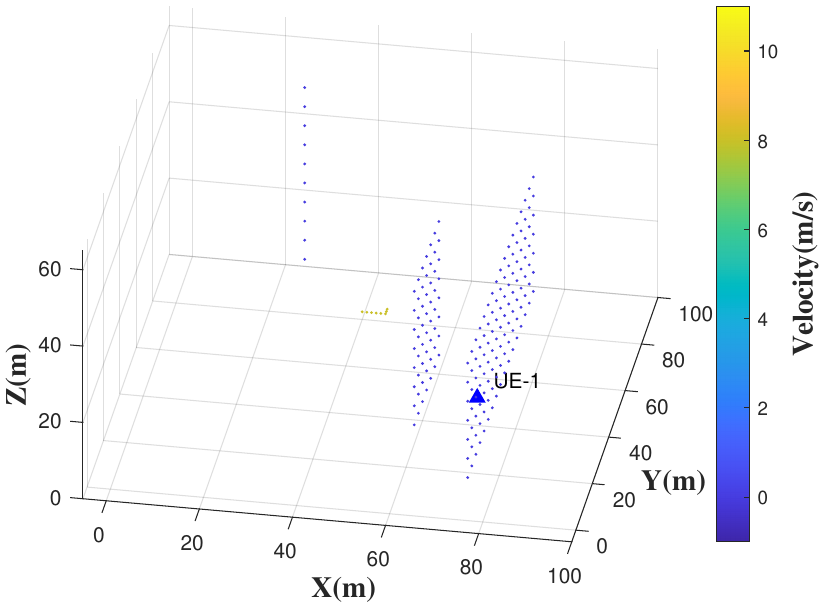}}
  	\subfigure[UL, Result, UE1 to BS2.]{
		\includegraphics[width=0.4\linewidth]{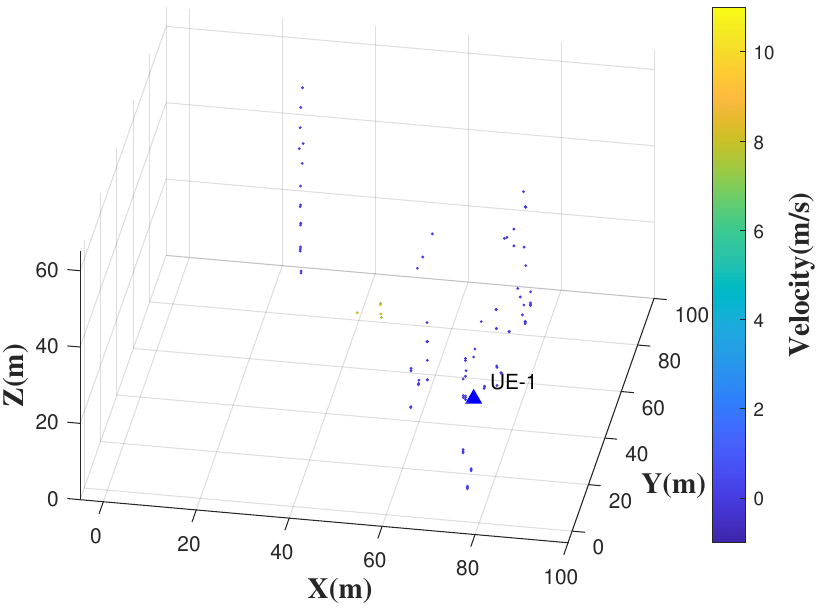}}
  \\
 	\subfigure[UL, Original, UAV1 to BS2.]{
		\includegraphics[width=0.4\linewidth]{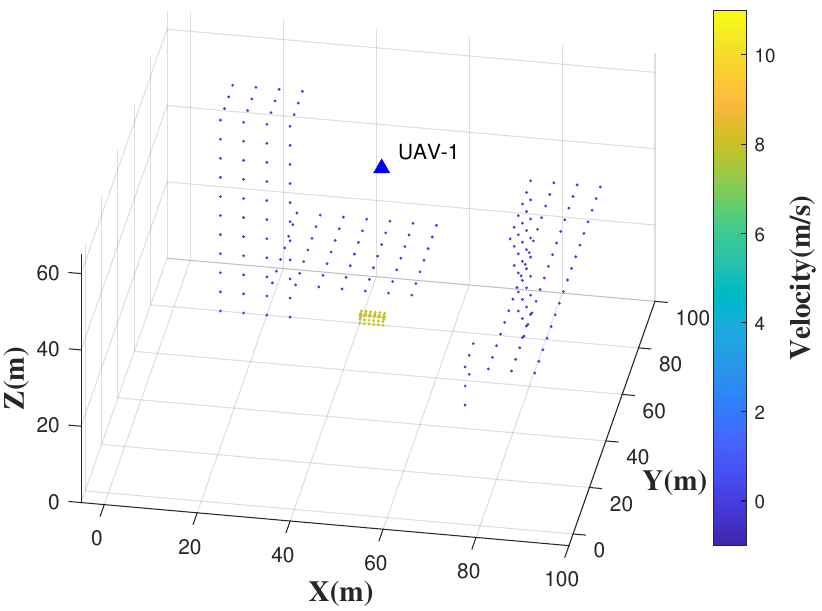}}
   	\subfigure[UL, Result, UAV1 to BS2.]{
		\includegraphics[width=0.4\linewidth]{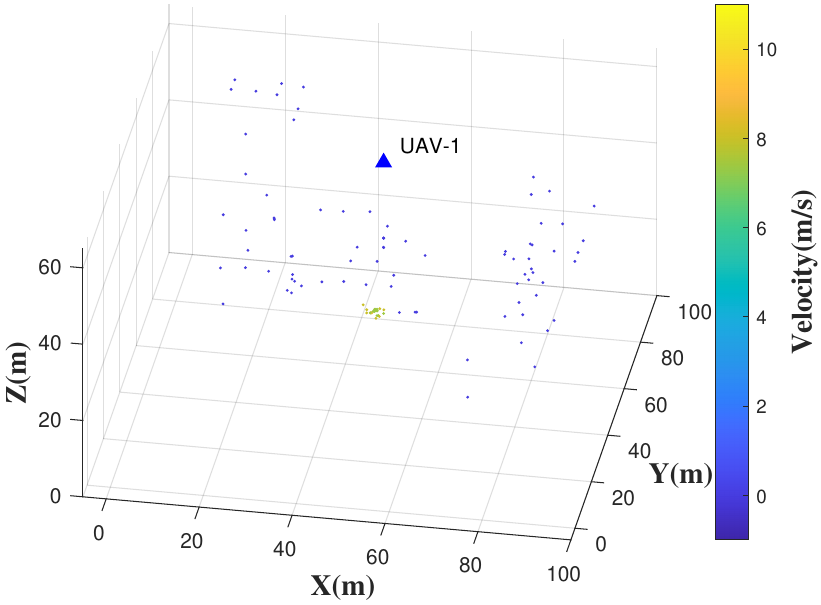}}
	\caption{Original points cloud and reconstruction result of a few sensing nodes in the DL and UL periods.}
	\label{fig:multi-view}
 \end{figure}
The environmental reconstruction results of some sensing nodes are shown in Fig.~\ref{fig:multi-view}, and the multi-node cooperative sensing environmental reconstruction results are shown in Fig.~\ref{fig:datalevel_fusion}. 

By comparing the single-node and multi-node sensing results, we can find that the multi-node cooperative ISAC 4D environmental reconstruction scheme can well overcome the problem of reconstructed point cloud disability caused by the single-node case due to the singularity of viewpoints.

\subsubsection{Comparison of Low-Density Point Cloud Enhancement Results}
\begin{figure*}[!htb]
	\centering
	\includegraphics[width=0.85\linewidth]{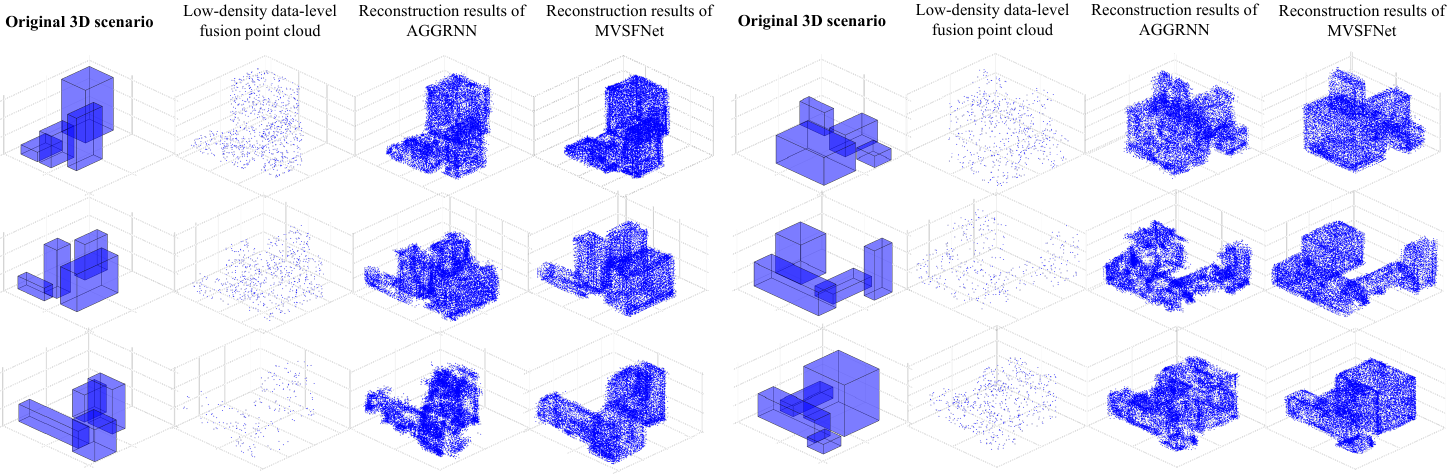}
	\caption{Selected images of original 3D scenario, low-density data-level fusion point cloud, AGGRNN reconstruction results, and MVSFNet reconstruction results.}
	\label{fig:resultallpic}
\end{figure*}


The enhancement performance for the selected randomly generated scenarios of the proposed low-density reconstruction point cloud enhancement network AGGRNN with MVSFNet is shown in Fig.~\ref{fig:resultallpic}. The simulation results show that, 
\begin{itemize}
\item Firstly, the results of the data-level fusion multi-node UL-DL collaborative ISAC 4D environmental reconstruction will lose a large amount of surface details, although it can basically depict the contours of the environment's scatterers. 
\item Second, the data-level fusion reconstructed point cloud enhanced by AGGRNN is significantly higher in density and can recover partial environmental surface details.
\item Finally, MVSFNet with multi-view feature-level fusion can recover more surface details compared with AGGRNN, and reconstruction deviation is smaller compared with the original scenario.
\end{itemize}

\subsection{Quantitative Analysis of Simulation Results}
\subsubsection{Comparative Experiments on High-Density Point Cloud Generation}

\begin{table}[!htb]
\footnotesize
    \centering
    \caption{Low-density reconstruction point cloud enhancement results on the generated dataset compared using F-Score@$1\%$ and CD.}
    \label{table:compare}
    \setlength{\tabcolsep}{0.7mm} 
    \begin{tabular}{@{}c|cc@{}}
    \toprule
    \textbf{Methods}                                      & \textbf{F-Score@$1\%$} & \textbf{CD}                        \\ \midrule
    AtlasNet\cite{groueix2018papier}                                     & 0.566       & 0.268                     \\
    PCN\cite{yuan2018pcn}                                          & 0.589       & 0.314                     \\
    FoldingNet\cite{yang2018foldingnet}                                   & 0.237       & 0.499                     \\
    TopNet\cite{tchapmi2019topnet}                                       & 0.414       & 0.336                     \\
    MSN\cite{liu2020morphing}                                          & 0.601       & 0.287                     \\
    GRNet\cite{xie2020grnet}                                        & 0.613       & 0.352                     \\
    \multicolumn{1}{l|}{\textbf{Data-level Fusion(Ours)}} & -           & \multicolumn{1}{l}{0.517} \\
    \textbf{AGGRNN(Ours)}                                 & 0.652       & 0.270                     \\
    \textbf{MVSFNet(Ours)}                                & \textbf{0.751}       & \textbf{0.191}                     \\ \bottomrule
    \end{tabular}
\end{table}
The quantitative results in Table \ref{table:compare} show that our proposed Multi-Modal model MVSFNet outperforms the other compared schemes in both Chmafer Distance and F-Score$@1\%$ metrics, and outperforms our proposed AGGRNN without supplemental local features.

\subsubsection{Ablation Study}
In order to demonstrate the superiority of our proposed model more completely, we will perform ablation study for transfer learning and fusion\&enhancement strategies.
\begin{itemize}
\item We perform an ablation study for transfer learning of MVSFNet, where the GFEM of MVSFNet loads a pre-trained weight file based on the ShapeNet dataset\cite{wu20153d} when using transfer learning, and conversely without using that pre-trained weights. The quantitative results in Table \ref{table:Transferlearning} show that transfer learning can optimize the performance of MVSFNet with other experimental conditions being the same.
\begin{table}[!htb]
\footnotesize
    \caption{Ablation Study (Transfer Learning)}
    \label{table:Transferlearning}
    \setlength{\tabcolsep}{0.7mm} 
    \centering
    \begin{tabular}{@{}c|cc@{}}
    \toprule
    \textbf{Transfer Learning} & \textbf{F-Score@$1\%$} & \textbf{CD} \\ \midrule
                               & 0.716               & 0.238       \\
        \checkmark                & \textbf{0.751}               & \textbf{0.191}       \\ \bottomrule
    \end{tabular}
\end{table}
\item We perform ablation study for Fusion\&Enhancement strategies, where MVSFNet cannot perform ablation study alone since the feature-level fusion strategy MVSFNet is based on the enhancement strategy AGGRNN. Quantitative results show in Table \ref{table:stra} that the ISAC environmental reconstruction performance is best when the multi-level fusion strategy is applied together with the low-density point cloud enhancement strategy, which proves the superiority and rationality of our proposed method. It is worth noting that direct low-density point cloud enhancement without applying a data-level fusion strategy can lead to deterioration of the environmental reconstruction performance, since the unfused point cloud accumulates weights to the corresponding 3D Grids, which in turn leads to disruption of the overall spatial structure.

\begin{table}[!htb]
    \centering
    \caption{Ablation Study (Fusion \& Enhancement)}
    \label{table:stra}
    \setlength{\tabcolsep}{0.7mm} 
    \begin{tabular}{@{}ccc|cc@{}}
    \toprule
    \multicolumn{3}{c|}{\textbf{Fusion\&Enhancement}}                                                                                                                                                                                  & \multirow{2}{*}{\textbf{F-Score@$1\%$}} & \multirow{2}{*}{\textbf{CD}} \\ \cmidrule(r){1-3}
    \textbf{\begin{tabular}[c]{@{}c@{}}Data-level\\ Fusion\end{tabular}} & \textbf{\begin{tabular}[c]{@{}c@{}}Enhancement\\ (AGGRNN)\end{tabular}} & \textbf{\begin{tabular}[c]{@{}c@{}}Feature-level\\ Fusion(MVSFNet)\end{tabular}} &                                      &                              \\ \midrule
    \checkmark                                                           &                                                                         &                                                                                  & -                                    & 0.517                        \\
    \checkmark                                                           & \checkmark                                                              &                                                                                  & 0.652                                & 0.270                        \\
                                                                         & \checkmark                                                              &                                                                                  & 0.545                                & 0.391                        \\
                                                                         & \checkmark                                                              & \checkmark                                                                       & 0.674                                & 0.332                        \\
    \checkmark                                                           & \checkmark                                                              & \checkmark                                                                       & \textbf{0.751}                       & \textbf{0.191}               \\ \bottomrule
    \end{tabular}
\end{table}

\end{itemize}

\section{Conclusion} \label{sec6}
We propose a deep learning based MNDUC ISAC 4D environmental reconstruction method in this research, which is based on the MUSIC algorithm in the echo processing part, and introduces multi-node UL and DL cooperation as well as multi-level fusion strategy, and optimizes the results of the environmental reconstruction by the deep learning models AGGRNN and MVSFNet. We have performed a large number of comparison and ablation studies for the proposed method, and the experimental results show that
\begin{itemize}
    \item The proposed multi-node cooperative sensing scheme combining the active and passive sensing can make up for the shortcomings of the single-node scheme's single viewpoint, and realize the comprehensive, multilevel, multifaceted environmental reconstruction results; 
    \item The proposed reconstruction points cloud density enhancement networks AGGRNN and MVSFNet can significantly reduce the gap between the multi-node cooperative ISAC 4D reconstruction results and the real scenario, and reduce the dependence of ISAC-based 4D environmental reconstruction methods on frequency bands and bandwidth resources.
\end{itemize}


\bibliographystyle{IEEEtran}
\bibliography{bare_jrnl_new_sample4}

\end{document}